\documentclass[printer]{aa}

\usepackage[utf8]{inputenc}
\usepackage{graphicx}
\usepackage{xcolor}
\usepackage[colorlinks=true,citecolor=blue,linkcolor=black,breaklinks=true]{hyperref}

\begin{document}

\title{Dust entrainment in photoevaporative winds:\\The impact of X-rays}

\author{R. Franz \inst{1}\thanks{rfranz@usm.lmu.de} \and G. Picogna \inst{1} \and B. Ercolano \inst{1,2} \and T. Birnstiel \inst{1,2}}

\institute{University Observatory, Faculty of Physics, Ludwig-Maximilians-Universit\"at M\"unchen, Scheinerstr.~1, 81679 Munich, Germany \and
    Excellence Cluster Origin and Structure of the Universe, Boltzmannstr.~2, 85748 Garching, Germany}

\date{Received 02 Sep 2019 / Accepted 08 Jan 2020}

\abstract
{X-ray- and EUV- (XEUV-) driven photoevaporative winds acting on protoplanetary disks around young T-Tauri stars may crucially impact disk evolution, affecting both gas and dust distributions.}
{We investigate the dust entrainment in XEUV-driven photoevaporative winds and compare our results to existing MHD and EUV-only models.}
{We used a 2D hydrodynamical gas model of a protoplanetary disk irradiated by both X-ray and EUV spectra from a central T-Tauri star to trace the motion of passive Lagrangian dust grains of various sizes.
The trajectories were modelled starting at the disk surface in order to investigate dust entrainment in the wind.}
{For an X-ray luminosity of $L_X = 2 \cdot 10^{30}\,\mathrm{erg/s}$ emitted by a $M_* = 0.7\,\mathrm{M}_\odot$ star, corresponding to a wind mass-loss rate of $\dot{M}_\mathrm{w} \simeq 2.6 \cdot 10^{-8} \,\mathrm{M_\odot/yr}$, we find dust entrainment for sizes $a_0 \lesssim 11\,\mu$m ($9\,\mu$m) from the inner 25\,AU (120\,AU).
This is an enhancement over dust entrainment in less vigorous EUV-driven winds with $\dot{M}_\mathrm{w} \simeq 10^{-10}\,\mathrm{M_\odot/yr}$.
Our numerical model also shows deviations of dust grain trajectories from the gas streamlines even for $\mu$m-sized particles.
In addition, we find a correlation between the size of the entrained grains and the maximum height they reach in the outflow.}
{X-ray-driven photoevaporative winds are expected to be dust-rich if small grains are present in the disk atmosphere.}


\maketitle


\section{Introduction}
\label{sec:Intro}

Planets form from the gas and dust surrounding newly born stars, whose physical properties and final dispersal are strongly influenced by the stellar irradiation from their host star.
In particular, high-energy radiation may warm up the disk atmosphere, launching a thermal wind \citep[see e.g.][]{Hollenbach-1994, Gorti-2009a, Alexander-2014}.
Models predict that this photoevaporative wind can ultimately disperse the disk and may have important consequences for the formation and evolution of planetary systems \citep{Alexander-2012a, Ercolano-2015a, Ercolano-2017b, Carrera-2017, Jennings-2018, Monsch-2019}.

Despite the potential influence of this process on the formation of planets, the magnitudes of photoevaporative winds are still largely uncertain, with model predictions diverging by several orders of magnitude \citep{Armitage-2011, Alexander-2014, Ercolano-2017a}.
One problem is that to date, the only direct evidence of these winds is blue-shifted forbidden-line emission towards T-Tauri stars, including [Ne\,II] 12.8$\mu$m and [O\,I] 6300\AA~\citep[e.g.][]{Hartigan-1995, Pascucci-2008, Rigliaco-2013, Natta-2014, Simon-2016, Banzatti-2019}.
While the intensity and low-resolution profiles of these lines can be matched very well by X-ray photoevaporation models \citep{Alexander-2008, Ercolano-2010, Ercolano-2016}, it has been demonstrated that these lines do not trace the base of the wind.
Furthermore, their extreme temperature dependence makes them a tracer of the heating mechanism of an already unbound wind rather than tracing the wind-driving mechanism itself \citep{Ercolano-2016}.
An additional problem is that high resolution data has revealed very complex line profiles which may include components emitted in a magnetically-driven wind \citep{Banzatti-2019}.

Different types of wind diagnostics would be desirable for constraining disk dispersal models; small dust grains -- which can be entrained by the wind -- may provide an interesting avenue towards this end \citep{Giacalone-2019}.
A previous work by \citet{Owen-2011a} has shown that grains up to about $2\,\mu$m in size (i.e. radius) can be lifted up and blown out by an EUV-driven wind around a Herbig Ae/Be star.
More recently, \citet{Hutchison-2016c} have investigated EUV-driven dust outflow by means of a two-fluid smoothed particle hydrodynamics (SPH) code \citep{Hutchison-2016b}; they find entrainment of grains of up to $4\,\mu$m around a $0.75\,$M$_\odot$ T-Tauri star (although they note that this value may drop to about $1\,\mu$m due to grains settling towards the disk midplane).
Both \citet{Owen-2011a} and \citet{Hutchison-2016c} show that the wind selectively entrains grains of different sizes from different radii.
This results in a dust population which spatially varies in the wind, due to the topology of the gas streamlines which propagate almost radially outwards.
At NIR wavelengths, this variable grain population produces a `wingnut' morphology which may already have been observed in the case of PDS 144N \citep{Perrin-2006}.
Yet \citet{Owen-2011a} could not reproduce the color gradient of the observations, which show redder emission at larger heights above the disk; they suggest that this could be because the synthetic observations might be dominated by emission from the smallest grains entrained in the flow.
Grain growth in the underlying disk \citep[see][]{Testi-2014}, which they neglected in their calculations for simplicity, could reduce the population of small grains, and may hence provide a solution to this color problem.
While it is unclear whether the observations of PDS 144N can be explained by dust entrainment in a photoevaporative wind, \citet{Owen-2011a} have demonstrated that a significant amount of small grains -- which dominate the opacity in the FUV -- do populate disk winds, and hence play an important role in their chemistry.

In this work, we study the entrainment of dust grains in an X-ray driven wind around a T-Tauri star.
For this we use a particle approach \citep{Picogna-2018}, bootstrapped onto a steady-state hydrodynamical simulation of a photoevaporating disk \citep{Picogna-2019}.
Our results aim to facilitate more detailed studies of the detectability of winds in scattered light, as well as wind opacity models.
The latter should allow for more realistic chemical modelling of the gas in the wind, enabling us to search for new wind diagnostics. 

This paper is organised as follows:
We present the numerical setup of the gas disk and photoevaporative disk wind, and the dust grain evolution in Section~\ref{sec:Methods}.
In Section~\ref{sec:Results}, we take a detailed look at what we can extract from the dust grain trajectories we obtain.
We discuss our findings in Section~\ref{sec:Discussion} and summarize them in Section~\ref{sec:Summary}.


\section{Methods}
\label{sec:Methods}

The dynamics of dust grains in protoplanetary disks can be studied either by directly integrating the orbits of a large number of dust `super-particles', which sample the local properties of the dust population, or by solving the collisionless Boltzmann equation for the particle distribution function.
For a population of very small (i.e. tightly coupled to the gas) dust particles, the Boltzmann equation can be reduced to the zero-pressure fluid equation \citep{Cuzzi-1993, Garaud-2004a}; this `two-fluid' approach has already been used to study planet-disk interactions \citep[e.g.][]{Paardekooper-2004, Paardekooper-2006, Zhu-2012}.
However, it is limited to a single population of small particles as it cannot account for the full velocity distribution of the grains at a single location, and it is not able to capture strong density gradients.

In contrast, a particle approach as implemented by \citet{Picogna-2018} has the notable advantage of following the evolution of solid particles with different physical properties, recovering the dust dynamics very well also in the limit where the grains are decoupled from the gas \citep{Youdin-2007a, Miniati-2010, Bai-2010b}.
This method has been successfully applied to the study of planet-disk interaction with both SPH and grid-based codes \citep{Fouchet-2007, Lyra-2009, Fouchet-2010, Ayliffe-2012, Zhu-2014}, and to modelling the draining of dust grains from the inner region of a photoevaporating transition disk \citep{Ercolano-2017b}.

\subsection{Gas disk with XEUV wind}

The set-up of the hydrodynamical calculations for the gas disk has been described in detail in \citet{Picogna-2019}, so here we limit ourselves to summarising the basic parameters of the specific run employed in this work.
We studied a protoplanetary disk of $M_\mathrm{disk} \simeq 0.01 \,M_*$ around a $M_* = 0.7 \,M_\odot$ T-Tauri star.
This star was set to emit X-ray and EUV radiation according to the emission line spectrum presented by \citet{Ercolano-2008b, Ercolano-2009}; the X-ray luminosity of the star was $L_X = 2 \cdot 10^{30}$\,erg/s, which is close to the median of the X-ray luminosity distribution for this stellar mass
\citep{Preibisch-2005}.

The hydrodynamics (HD) simulations were performed via a modified version of the \texttt{Pluto} code. In this version, at each HD step, temperatures are updated according to the local ionization parameter\footnote{The ionization parameter is defined as $\xi = L_* / (n\,r^2)$, with $L_*$ the stellar luminosity, $n$ the gas density, and $r$ the (spherical) radial distance from the star.} and column density to the central source \citep[for further details, see][]{Picogna-2019}.
The temperature parametrization was obtained via detailed radiative-transfer calculations using the \texttt{Mocassin} code \citep{Ercolano-2003, Ercolano-2005, Ercolano-2008a}.
Within the \texttt{Pluto} code used for the gas evolution \citep{Mignone-2007}, we employed a 2.5D Eulerian grid in spherical coordinates.\footnote{`2.5D' meaning a 2D coordinate system $(r,\vartheta)$ with 3D velocity information $(v_r, v_\vartheta, v_\varphi)$.}
To avoid any boundary effects, a large radial range of $0.33 \leq r\,[\mathrm{AU}]\leq 1000$ in 412 logarithmically-spaced steps was modelled, with $0.005 \leq \vartheta \leq \pi/2$ in 320 uniform steps.

\begin{figure}
    \centering
    \includegraphics[scale=0.55]{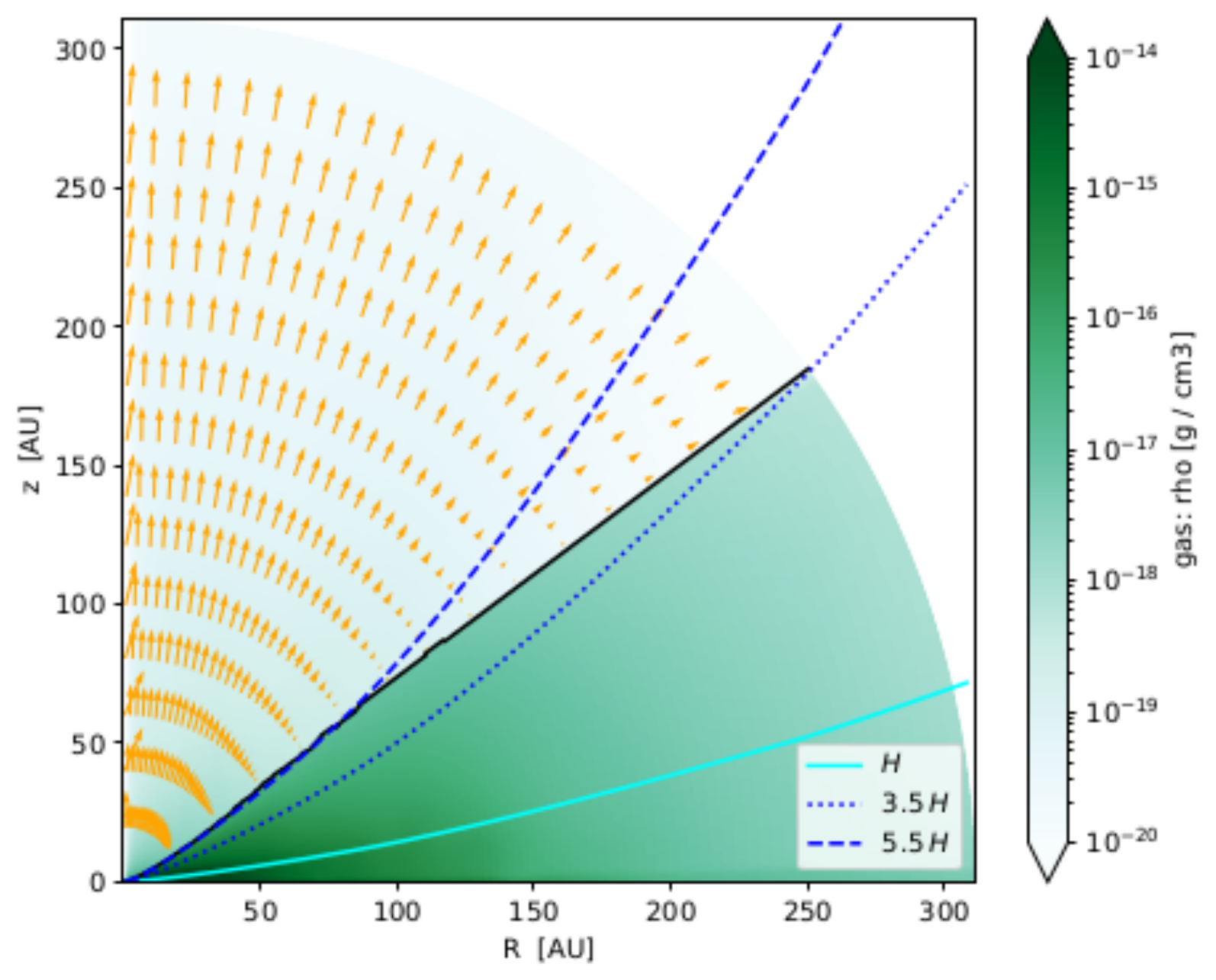}
    \caption{Density (green) and velocity map (yellow arrows: $\vec{v}$) of the gas disk model from \citet{Picogna-2019}.
    The base of the XEUV flow (i.e. the location of the largest gradient in temperature) is shown in black, and coincides with a strong drop in density. $v_r$ is pointing radially outwards everywhere in the wind but very close to the sonic surface where $\vec{v}$ points outwards, but away from the disk (for $R < 200\,$AU).
    The cyan line represents $z(R) = H$; additional lines for $3.5\,H$ (dotted blue) and $5.5\,H$ (dashed blue) show the range of scale heights the disk surface covers.}
    \label{fig:gasdisk}
\end{figure}

The $v_r$- and $\varrho$-profiles of the gas disk employed for the simulations are showcased in Fig.~\ref{fig:gasdisk}; we used only the inner $\approx 300\,$AU (352 cells) of the hydrodynamical grid in order to follow the dust evolution because the mass loss due to the photoevaporative wind becomes negligible at larger radii \citep[see][]{Picogna-2019}.

The (sonic) disk surface is defined as the plane where the gas velocities change from locally sub- to super-sonic.
The base of the X-ray and EUV (XEUV) flow, by contrast, is given by the location of the largest temperature gradient \citep[see e.g.][]{Ercolano-2009}, and lies slightly below the sonic surface.
It coincides with a strong drop in gas density (see Fig.~\ref{fig:gasdisk}).

Due to the high grid resolution employed and the disk having settled into a stable (quasi-equilibrium) state, the (numerically computed) base of the wind is quite smooth \citep[as would be expected; see also][]{Bai-2017}; hence, we forgo additional artificial smoothing.
In Fig.~\ref{fig:gasdisk}, we have added lines indicating one, 3.5, and 5.5 scale heights $H$, the latter two framing the base of the flow; here we use $H = h \cdot R$, and $H = c_s / \Omega_K$, with $c_s$ the local sound speed and $\Omega_K = \sqrt{G \, M_* / r^3}$ the Keplerian orbital velocity.

The flow base being located at a scale height of $h \simeq 0.08$ (0.14; 0.23) at $R = 10\,$AU (100\,AU; 300\,AU) implies that the disk is still rather hot and puffed up and that the dust grains have to travel rather far above the midplane if they are to enter the wind region \citep[see comparable simulations by][]{Owen-2012a, Bai-2016a}.

\subsection{Dust grains}

The dust grains were modelled as passive Lagrangian particles inserted in the steady-state gas solution, as originally implemented by \citet{Picogna-2018}, to whom we refer for the details of the implementation.\footnote{We are dropping their non-inertial $\vec{F}_\mathrm{nonin}$ because we are not including a planet in our simulations.}
The motion of these particles is driven by their gravitational attraction towards the central star, the drag force from the surrounding gas, and turbulent diffusion below the disk surface as prescribed by \citet{Charnoz-2011}.

\citet{Ormel-2018} give a concise comparison of their stochastic equation of motion to the strong-coupling approximation of \citet{Charnoz-2011}.
Although the former may be preferable for modelling grain motion within the disk, we are mainly interested in what happens once a grain enters the wind region where the gas density and thus turbulence are low; hence, just like \citet{Giacalone-2019} proceed for their MHD-wind model of dust motion, we do not optimize our model for the disk interior.

Above the disk surface, gas densities are too low to induce kicks (see Fig.~\ref{fig:gasdisk}), allowing us to neglect an otherwise necessary \citep{Flock-2017}, more intricate distinction between MRI and VSI.\footnote{\citet{Flock-2017} investigate this difference and conclude that the VSI may be more adept at lifting up grains.}

\subsubsection{Grain sizes}

In their EUV-only simulations, both \citet{Owen-2011a} and \cite{Hutchison-2016c} find that grains around $\mu$m size are entrained.
In the MRI computations of \citet{Miyake-2016}, the grain distribution considered is $0.1 \leq a_0\,[\mu\mathrm{m}] \leq 100$, and the maximum entrainable grain size is found to decrease very steeply with the (cylindrical) midplane radius $R = \sqrt{x^2+y^2} \,.$
Furthermore, \citet{Giacalone-2019} investigate $5 \cdot 10^{-3} \lesssim a_0\,[\mu\mathrm{m}] \lesssim 5$.

Therefore, we ran an initial set of simulations with grain sizes $10^{-3} \leq a_0\,[\mu\mathrm{m}] \leq 10^{2}$, which established that for our model, the size barrier for wind blow-out lies between 5 and $15\,\mu$m.
On the basis of this initial experiment, we restricted our size range to $0.01 \leq a_0\,[\mu\mathrm{m}] \leq 20$, with steps of $\Delta a_0 = 1\,\mu$m for $1 \leq a_0\,[\mu\mathrm{m}] \leq 15$.
We forwent a higher size resolution in favor of increasing spatial resolution, that is tracing more dust grains per each size.
All different $a_0$ modelled are listed in Table~\ref{tab:counts} (and Fig.~\ref{fig:dust-positions}); per size, we simulated the trajectories of at least 5,000 dust grains, yielding at least 25 grains per 1\,AU of launching radius along the disk surface.

\subsubsection{Internal grain density}

Following \citet{Owen-2011a} and in order to facilitate a direct comparison to their results, we assume a uniform internal density of $\varrho_\mathrm{grain} = 1$\,g/cm$^3$ for the dust particles.
This value is on the lower end of the $0.3 \lesssim \varrho_\mathrm{grain}\,[\mathrm{g/cm^3}] \lesssim 6.2$ interval established by \citet{Love-1994}, and agrees best with the values \citet{Joswiak-2007} find for material of cometary origin ($0.6 \lesssim \varrho_\mathrm{grain}\,[\mathrm{g/cm^3}] \lesssim 1.7$).
Similar values are used in other works, too \citep[e.g.][]{Tamfal-2018, Owen-2019}.

Other models employ somewhat different values for $\varrho_\mathrm{grain}$.
For instance, \citet{Li-1997} and \citet{Miyake-2016} use the average value given by \citet{Love-1994}, $\langle \varrho_\mathrm{grain} \rangle \simeq 2$\,g/cm$^3$; \citet{Hutchison-2016c} and \citet{Flock-2017} employ $\varrho_\mathrm{grain} = 3$\,g/cm$^3$, and \citet{Weingartner-2001} and \citet{Giacalone-2019} use $\varrho_\mathrm{grain} = 3.5$\,g/cm$^3$.
These values are closer to the ones \citet{Joswiak-2007} find for asteroidal material, which may have been heated slightly less than the cometary grains.
Future on-site analysis of interplanetary and interstellar dust grains will provide further constraints on these intervals \citep[e.g.][Destiny+]{Arai-2018}.

\subsubsection{Initial positioning}

We position our grains directly on the base of the flow which is located slightly below the disk surface.
This allows us to study their trajectories from when they enter the wind-dominated region above the disk.
Within $0.33 \leq r\,\mathrm{[AU]} \leq 200$, we use a random distribution uniform in $r$ for the initial placement.

\begin{figure*}
    \centering
    \includegraphics[scale=0.55]{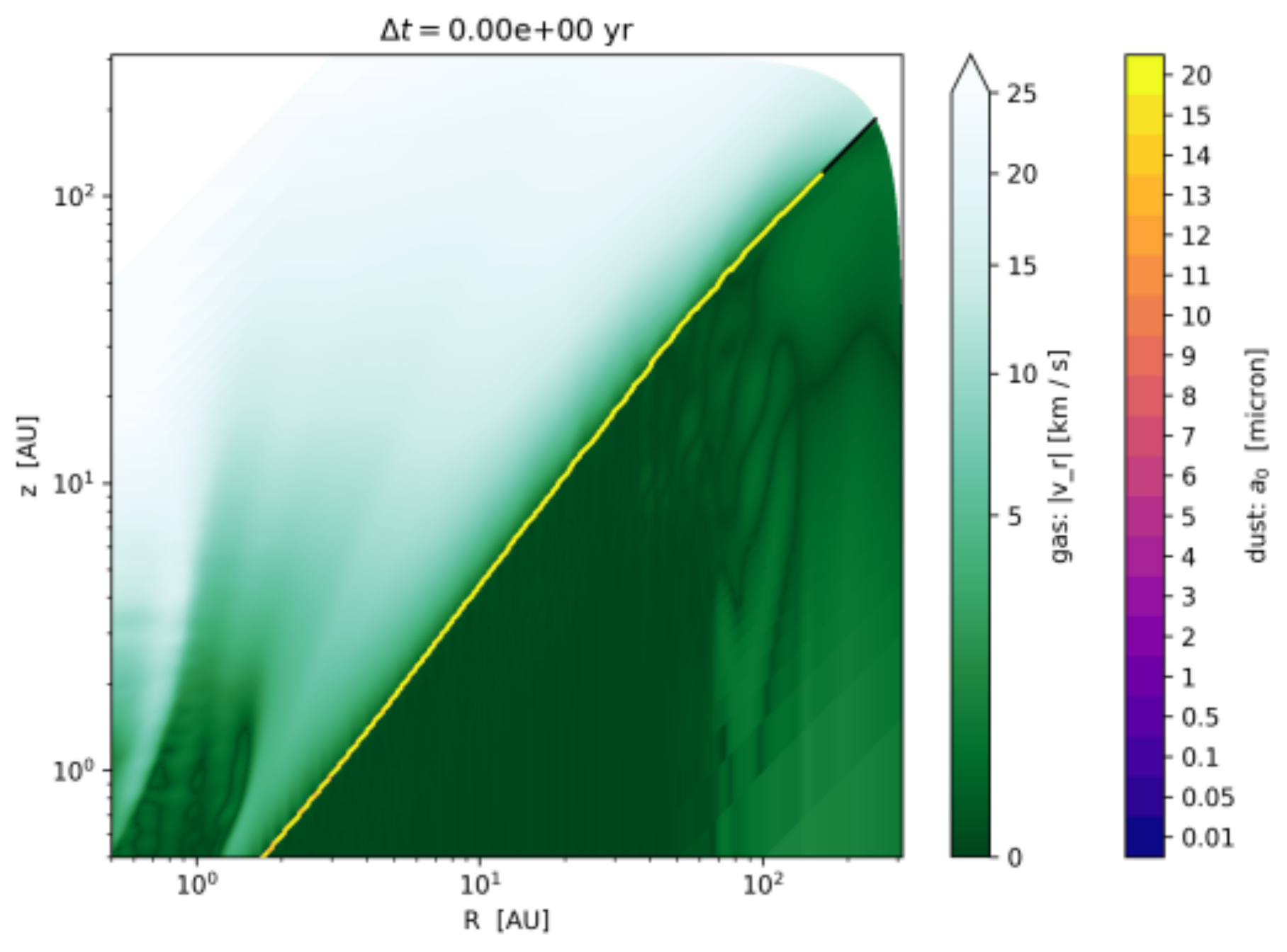}
    \includegraphics[scale=0.55]{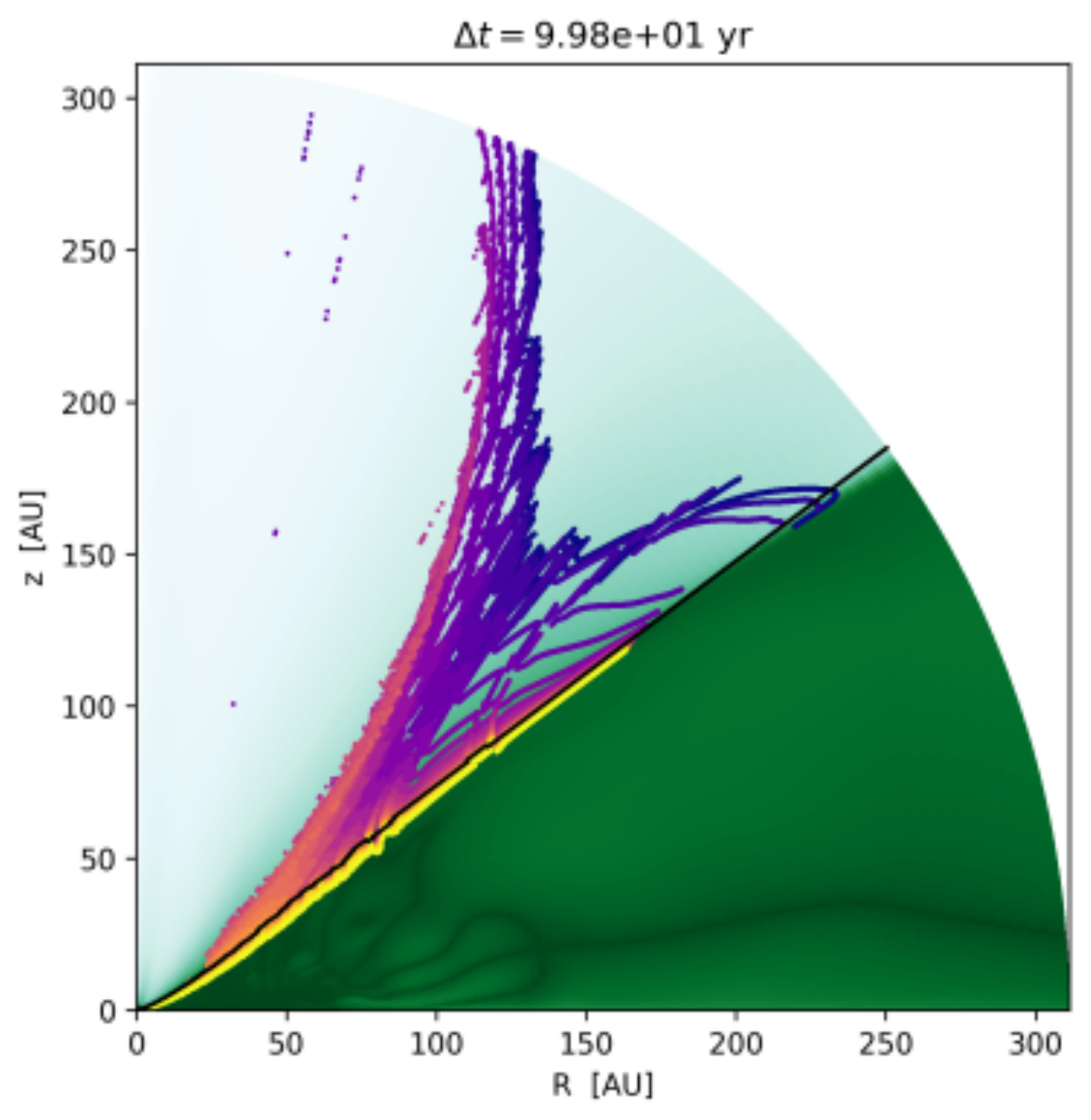}
    \caption{\textit{Left}: Initial placement of the dust grains at $\Delta t=0$ at the base of the photoevaporative flow (black), slightly below the sonic surface.
    The grains are colored according to their size and the color scale (scaling with $\sqrt{|v_r|}$) of the gas disk represents the extent of its local radial velocity (with $-2 \lesssim v_r \,\mathrm{[km/s]} \lesssim 30\,$).
    The gas map is mostly smooth in the region of interest, that is everywhere but close to the midplane at high $R$.
    \textit{Right}: A snapshot of the grain positions at $\Delta t \simeq 100\,$yr, all else equal.
    The very low spread of the lines of individual grain sizes is due to the initial setup, placing particles directly on the disk surface without a spread in their initial velocities. For comments on this, see also Section~\ref{sec:Results} and Appendix~\ref{sec:app:vel-sigma}.}
    \label{fig:dust-positions}
\end{figure*}

The left panel of Fig.~\ref{fig:dust-positions} shows the initial grain positioning along the base of the flow (in black), with the dust grains colored according to their size.

Our model is intended to be combined with a vertical mixing prescription later on in order to extract a realistic dust density distribution in the wind.
So we may very well, just as \citet{Hutchison-2016c} note they did, model grain sizes that will not migrate far enough vertically to actually reach the wind region \citep[see also][]{Youdin-2007b, Krijt-2016}.

The dust grains were placed -- and remain -- well outside of the sublimation radius applicable for a $M_* = 0.7\,\mathrm{M}_\odot$ star \citep{Giacalone-2019}.
However, we note that the intense X-ray radiation from the young stellar object may destroy PAH-size grains in the disk atmosphere before they can be entrained in the wind \citep{Siebenmorgen-2010, Siebenmorgen-2012}; we did not include such events in our simulation.

\subsubsection{Initial velocities}

We initialised our dust particles to start from a quasi-equilibrium; so we set both $v_{r,0}$ and $v_{\vartheta,0}$ to 0 because the local gas velocities are quite low anyways and will hence not cause a strong upwards motion ($\left| v_\vartheta \right| \lesssim 50\,$m/s along the disk surface compared to $\left| v_r \right| \lesssim 200\,$m/s).

For $v_{\varphi,0}$, we assumed a Keplerian velocity of $v_{\varphi,0} = r \, \Omega_K$; since the starting positions are at $z \gtrsim 3.5\,H$ (see Fig.~\ref{fig:gasdisk}), we computed the Keplerian speed for the spherical radius $r$, and not for the midplane radius $R$.

\subsubsection{Further limitations}

To cut computational costs and allow for a reasonable amount of particles to be modelled, we made a series of simplifying assumptions:

Firstly, we neglected MHD effects.\footnote{For an in-depth treatment of dust in a magneto-centrifugal disk wind with a setup similar to ours, see \citet{Giacalone-2019}.}

Secondly, we did not include self-gravity from the disk. We show in Appendix~\ref{sec:app:disk-gravity} that this simplification should not significantly affect our results.

Thirdly, we did not include dust-gas back reactions.
As shown by \citet{Dipierro-2018} and \citet{Tamfal-2018}, these are important in the disk midplane; but we focus our modelling efforts on the wind regions above the disk, where dust-to-gas ratios are not expected to be enhanced \citep{Krijt-2016}.
    
Fourthly, dust-dust interactions were neglected.
The gas drag accelerates the dust grains to at least a few km/s (i.e. $v_{r,\mathrm{esc}} = \sqrt{2\,G\,M / r} \approx 11\,\mathrm{km/s}$ at $r = 10\,$AU, or $v_{r,\mathrm{esc}} \approx 2\,\mathrm{km/s}$ at $r = 300\,$AU), but the dust densities in the wind are much lower than around the disk midplane.
At the latter, the growth time scale is already around $10^2$ to $10^3\,$yrs \citep{Birnstiel-2016}.
We shall see below that therefore, assuming no interactions provides a reasonable approximation.


\section{Results}
\label{sec:Results}

We traced the trajectories of the dust grains over time until they either leave the computational domain or until the simulation time frame of $\Delta t_\mathrm{sim} \simeq 2.2\,$kyr ends.

If they left the domain, they were replaced by a new grain of the same size, placed as described in Section~\ref{sec:Methods}; the actual amount of dust particles modelled per $a_0$ is listed in Table~\ref{tab:counts}.\footnote{Since some of the large particles may leave the computational domain due to the gas motion below the disk surface dragging them out, more than the minimum of 5,000 (see Section~\ref{sec:Methods}) trajectories are modelled for all grain sizes.}
Merely for visualizing the actual simulation, the right panel of Fig.~\ref{fig:dust-positions} shows a snapshot of the simulation after around 100\,yr.

\begin{table}
    \centering
    \caption{Number of modelled trajectories per $a_0$.}
    \label{tab:counts}
    \begin{tabular}{*{6}{r}}
        \multicolumn{1}{l}{$a_0$[$\mu$m]} & \multicolumn{1}{l}{N} & \multicolumn{1}{l}{$a_0$[$\mu$m]} & \multicolumn{1}{l}{N} & \multicolumn{1}{l}{$a_0$[$\mu$m]} & \multicolumn{1}{l}{N} \\ \hline
        0.01 & 82106  &  4 & 27165  &  11 & 5126 \\
        0.05 & 82293  &  5 & 21606  &  12 & 5064 \\
        0.1 & 80834  &  6 & 16866  &  13 & 5054 \\
        0.5 & 67418  &  7 & 12367  &  14 & 5072 \\
        1 & 57766  &  8 & 8996  &  15 & 5063 \\
        2 & 46456  &  9 & 7357  &  20 & 5056 \\
        3 & 35644  &  10 & 6356  &  (total) & 583665 \\
    \end{tabular}
\end{table}

\begin{figure*}
    \centering
    \includegraphics[scale=0.385]{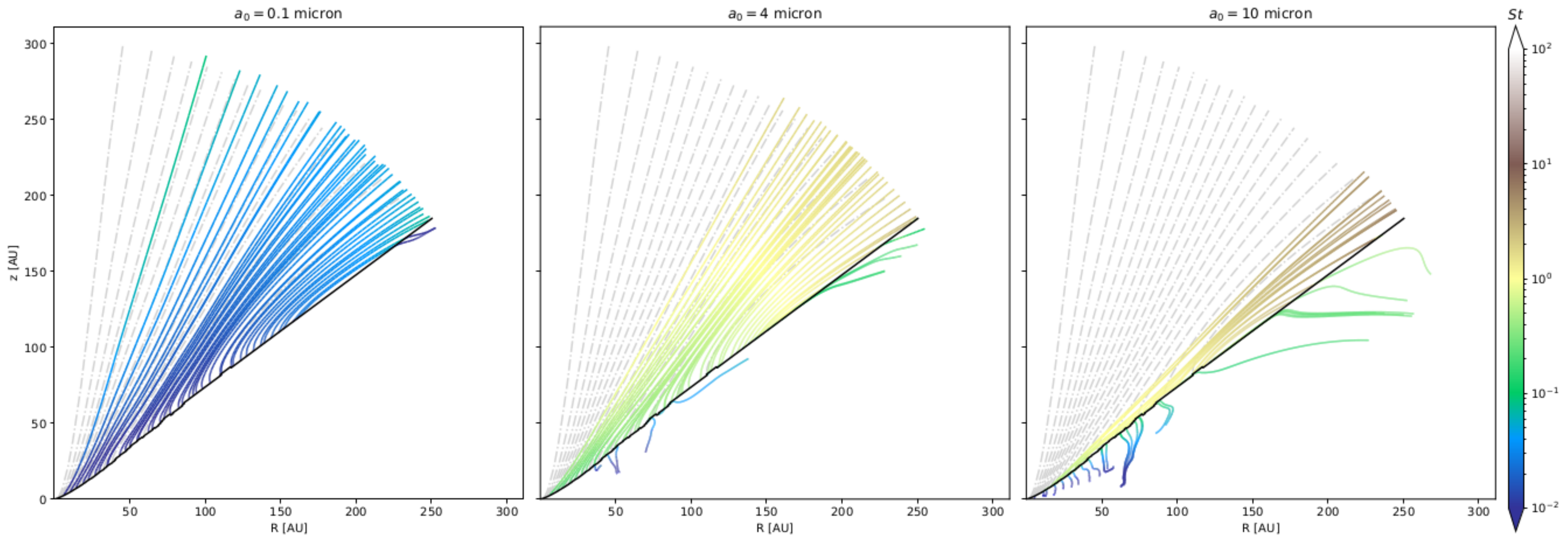}
    \caption{Randomly selected dust trajectories for $a_0$ = 0.1, 4, and 10$\,\mu$m (left, middle, and right panels, respectively).
    The trajectory color represents the local value of the Stokes number.
    Entrained dust grains, launching from the launching region (black), migrate upwards on the colorbar the as they move to regions of lower gas density.
    Gas streamlines are shown in dash-dotted grey.
    (For all simulated grain sizes, see Fig.~\ref{fig:trajectories-samples-all}.)}
    \label{fig:trajectories-samples-3}
\end{figure*}

A selection of trajectories obtained from the simulation is shown in Fig.~\ref{fig:trajectories-samples-3}; for clarity, we limit ourselves to plots for three distinguished grain sizes (0.1, 4, and 10$\,\mu$m) below.
These were chosen because they represent the three major varieties of grains encountered (see Appendix~\ref{sec:app:plots-all}).
Panels containing the complete set of 20 different $a_0$ are included in Appendix~\ref{sec:app:plots-all}.

In general terms, the grains analyzed are either fully entrained (blown out by the XEUV wind, leaving the computational domain above the disk surface), fall back below the base of the flow at $R \gtrsim 160\,$AU, or are not even picked up by the wind despite the turbulent kicks allowing for upwards motion.
Additionally, we find that trajectories for a given grain size do (almost) never intersect, wherefore different starting positions will lead to different paths in the wind.
Thus, the initial positioning of a grain of size $a_0$ along the launching region pre-determines which wind regions it can populate.
This matches with -- and is a direct result of -- the gas wind velocity map seen in Fig.~\ref{fig:gasdisk} pointing radially outwards almost everywhere.

\subsection{Robustness of the initial velocity setup}

In Section~\ref{sec:Methods}, we have described a rather simplistic initial velocity setup, with $\vec{v}_0$ depending only on $\Omega_K$.
However, in order for our grains to even reach the 3.5...5.5 $H$ which we launch them from (see Fig.~\ref{fig:gasdisk}), the gas must have some degree of turbulence, implying some variety in the initial velocities.

To account for this, we have run a series of tests with a Gaussian spread of $\sigma(v_i) = 100$\,m/s in all three directions $i \in \lbrace r; \vartheta; \varphi \rbrace$ of the initial velocity vector $\vec{v}_0$.
This value was chosen because it is an overestimate of the fragmentation speeds given by \citet[][10\,m/s]{Birnstiel-2009} and \citet[][$\lesssim 8\,$m/s for silicates, $\lesssim 80\,$m/s for icy aggregates]{Wada-2013}, and because it is slightly higher than the upwards speed of the gas which we find along the base of the flow, $|v_\vartheta| \lesssim 50\,$m/s.
So it should make for a suitable approximation of a velocity spread introduced by turbulent vertical mixing.

The results obtained do not deviate significantly from those retrieved without this spread; therefore we will proceed to show only the latter. For a more extensive elaboration on the similarities and differences identified, see Appendix~\ref{sec:app:vel-sigma}.

\subsection{Dust coupling to the gas}

In Fig.~\ref{fig:trajectories-samples-3}, the trajectories are colored by their Stokes number $St = t_\mathrm{stop} \cdot \Omega_K$, with $t_\mathrm{stop} = m_\mathrm{dust} \, v_\mathrm{dust} / F_\mathrm{drag}$ their local stopping time.\footnote{For the definition of the drag force $F_\mathrm{drag}$ employed here, see \citet{Picogna-2018}.}
Besides, gas streamlines spaced by 5\% of the total mass-loss rate of the gas in the wind region (i.e. $\dot{M}_\mathrm{w}$) have been added in (dash-dotted) grey for direct comparison.

$St \ll 1$ indicates that the dust motion is well-coupled to the gas flow; hence $St \ll 1$ is needed for a particle to be lifted up by the wind, since it is only affected by gas drag and stellar gravity.
Indeed, we find all entrained grains to have $St \lesssim 0.4$ when they are picked up by the wind (i.e. colors from blue to green).
While in the wind, they are sped up by the rather fast photoevaporative flow (of up to $v_r \lesssim 30\,$km/s, see Fig.~\ref{fig:dust-positions}), which leads to a steady increase in their speed, and thus also $St$; the latter may grow by up to an order of magnitude.

At low $St$, the dust grains follow the gas flow (see especially the left panel of Fig.~\ref{fig:trajectories-samples-3}); at $St \rightarrow 1$ however, they decouple from the gas flow (see the middle and right panels of Fig.~\ref{fig:trajectories-samples-3}, especially for higher $R$).
For $R \gtrsim 160\,$AU, the gas streamlines -- in particular those close to the disk -- start to bend towards it because the stellar irradiation is starting to decline this far out.
As a result, the dust grains that have already reached a relatively high radial velocity at this point overshoot the gas streamlines.
Further inwards and at higher $z$, the dust trajectories fall below the gas streamlines if they become decoupled.

\begin{figure*}
    \centering
    \includegraphics[scale=0.56]{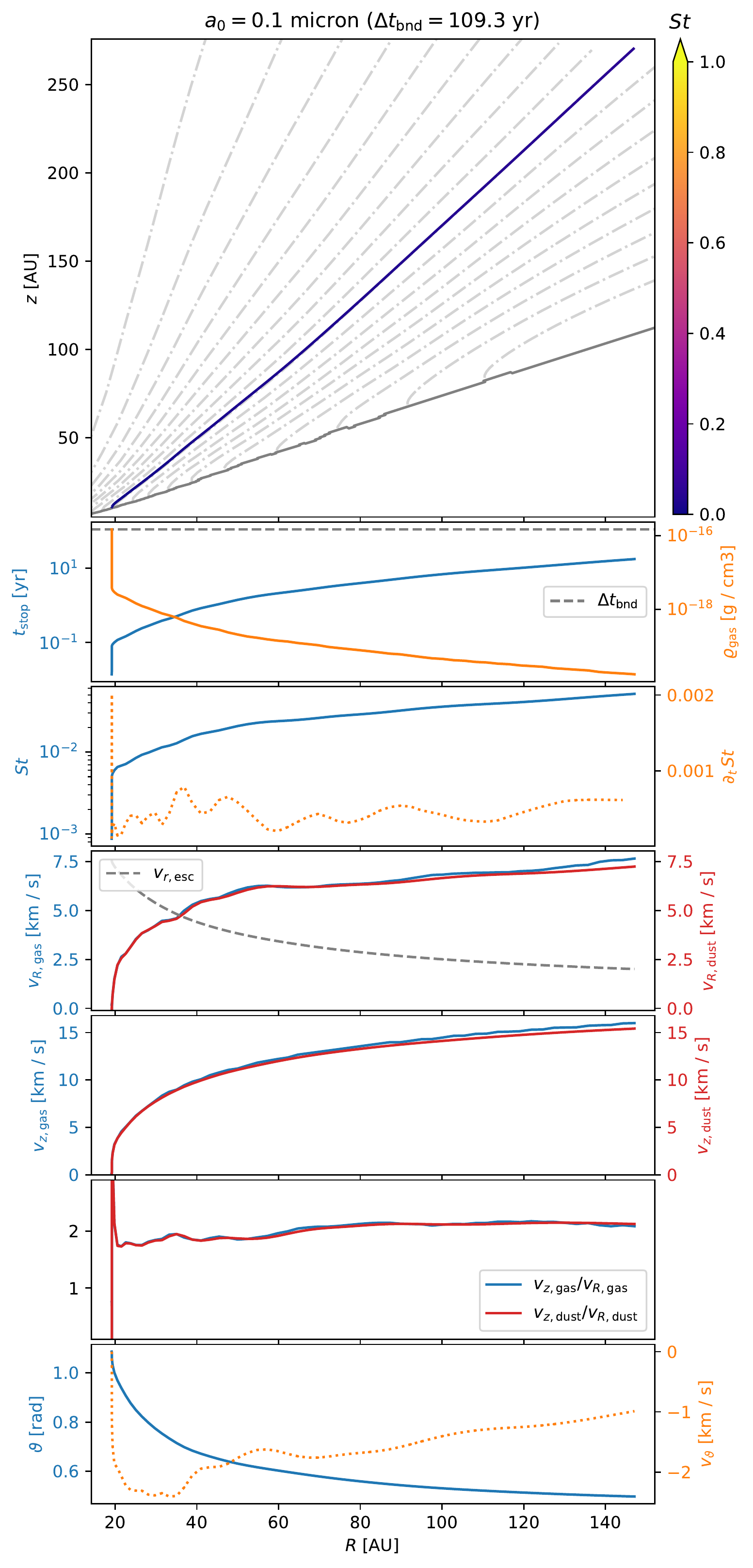}
    \includegraphics[scale=0.56]{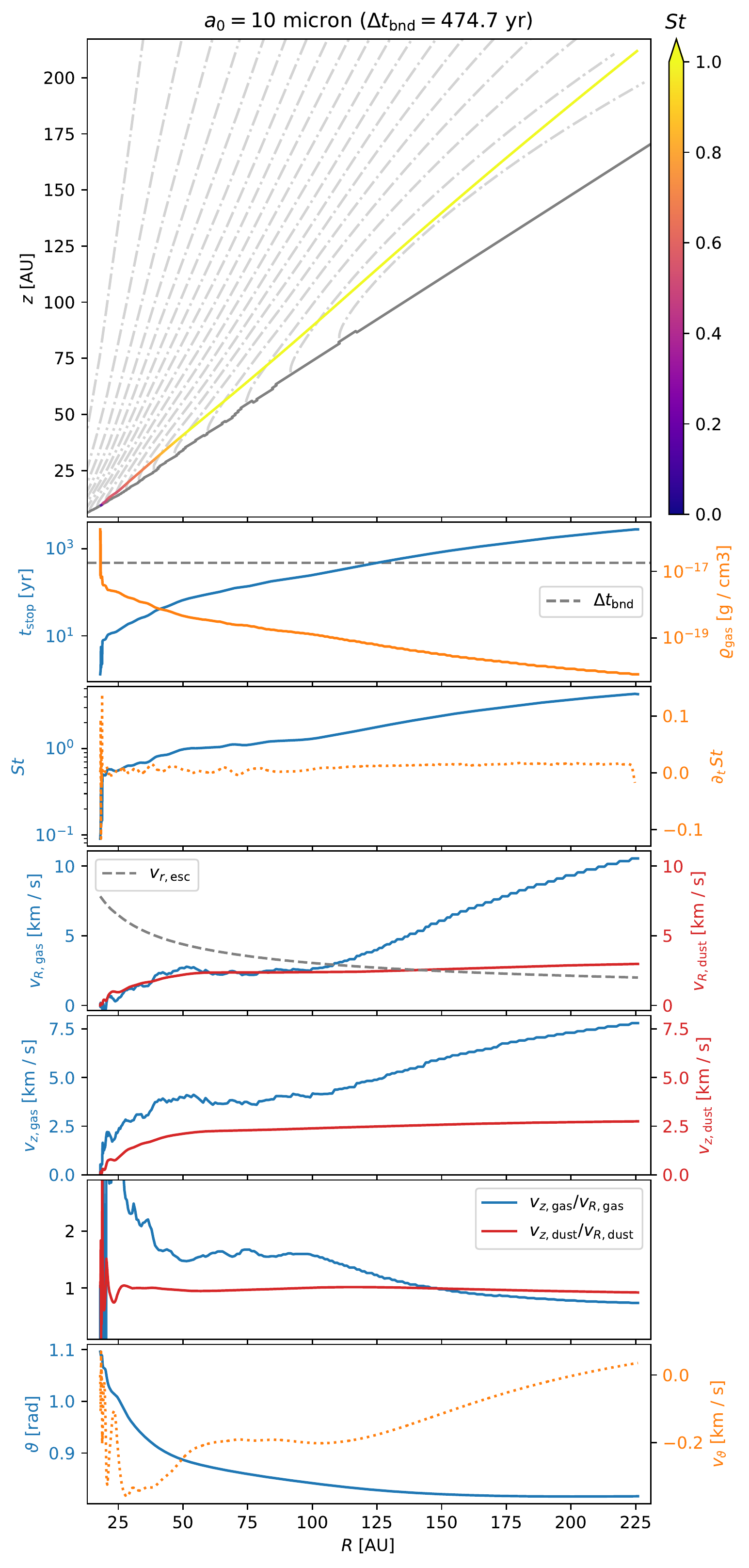}
    \caption{Analysis of two dust trajectories (left: $a_0=0.1\,\mu$m, right: $a_0=10\,\mu$m) entrained in the photoevaporative outflow from $R \simeq 20\,$AU.
    The top panels show the actual motion in the $(R,z)$-plane colored by $St$.
    The lower panels illustrate, from top to bottom, $t_\mathrm{stop}$ and $\varrho_\mathrm{gas}$, $St$ and $\partial_t \, St$, the horizontal speed $v_R$ of gas and dust, the vertical speed $v_z$ of gas and dust, a comparison of the direction of the motion $v_z/v_x$ for gas and dust, and $\vartheta$ and $v_\vartheta$.
    See the text for an in-depth commentary.}
    \label{fig:trajectories-analysis}
\end{figure*}

For an in-depth analysis of the dust motion, Fig.~\ref{fig:trajectories-analysis} shows two randomly selected dust particles launched from $R \simeq 20\,$AU.
These are representative of the dust grains picked up by the wind from this $R$, with other grains entrained from around this $R$ showing very similar trajectories; we opted for a launching point rather close to the star in order to showcase fully-entrained grains.\footnote{As we will see further down, this is also the $R$ from which the most massive grains are entrained.}

The $0.1\,\mu$m particle (left column) remains entrained in the photoevaporative flow and follows the gas motion almost perfectly; its Stokes number remains small ($St < 0.1$) throughout its trajectory.
Its $t_\mathrm{stop}$ also stays small, even after the strong increase (of a factor of about 4) it experiences when being picked up by the wind, simultaneous to the strong decrease in the density of the surrounding gas (of a factor of almost 100).
While the grain is at $r < 300\,$AU, $t_\mathrm{stop}$ is always smaller than the time needed for blow-out to 300\,AU (dashed grey line in the second panel), which may serve as a further indication that the particle stays coupled to the gas.
The fourth through sixth panels show a comparison of gas (blue) and dust (red) velocities for $v_R$ and $v_z$, where we can also observe a strong coupling.
It is only at $r \rightarrow 300\,$AU that a slight deviation of $\vec{v}_\mathrm{dust}$ from $\vec{v}_\mathrm{gas}$ occurs; the decoupling does not necessarily coincide with the particles reaching escape velocities (dashed grey lines in the fourth panel).
This means that even sub-micron particles start decoupling from the gas flow at high $r$, that is after picking up enough momentum from the wind. Also, with the curve for $\vartheta$ flattening down and $v_\vartheta$ being rather small in comparison to $v_R$, the motion of the showcased particle is almost fully radial at larger $R$. As $v_\vartheta < 0$, a small additional upwards component remains.

On the other hand, the $10\,\mu$m particle (right column) decouples from the gas flow within the first few AU of entering the wind region; simultaneously, its Stokes number quickly becomes $St > 0.5$.
Its $t_\mathrm{stop}$ grows even larger than the time passing between wind pick-up and leaving the domain (again, dashed blue line).
The XEUV wind drags the grain along, increasing its $v_R$ which remains comparable to $v_{R,\mathrm{gas}}$ for the first part of its trajectory.
By contrast, $v_{z}$ quickly diverges from the gas flow, which leads to the direction of the dust motion clearly (visually) differing from the gas streamlines, intersecting multiple ones.
At $R \gtrsim 160$\,AU, the gas flow starts pointing back down towards the disk, but since the grain is already decoupled, it does not seem to be affected by this; in the sixth panel, we see that its direction of motion remains almost constant after the initial acceleration, as would be expected for high $t_\mathrm{stop}$.

Despite the gravitational pull acting on the $10\,\mu$m grain, its $v_R$ and $v_z$ are slightly increasing for $R \gtrsim 100\,$AU.
This is caused by the relatively high difference in gas and dust velocity $\left| v_\mathrm{gas} - v_\mathrm{dust}\right|$; even if this additional speed-up were missing, the grain would still reach $v_{r,\mathrm{esc}}$ well within $r < 300\,$AU, as the grey dashed line in the fifth panel shows.

\subsection{Dust timescales in the wind}

\begin{figure*}
    \centering
    \includegraphics[scale=0.4]{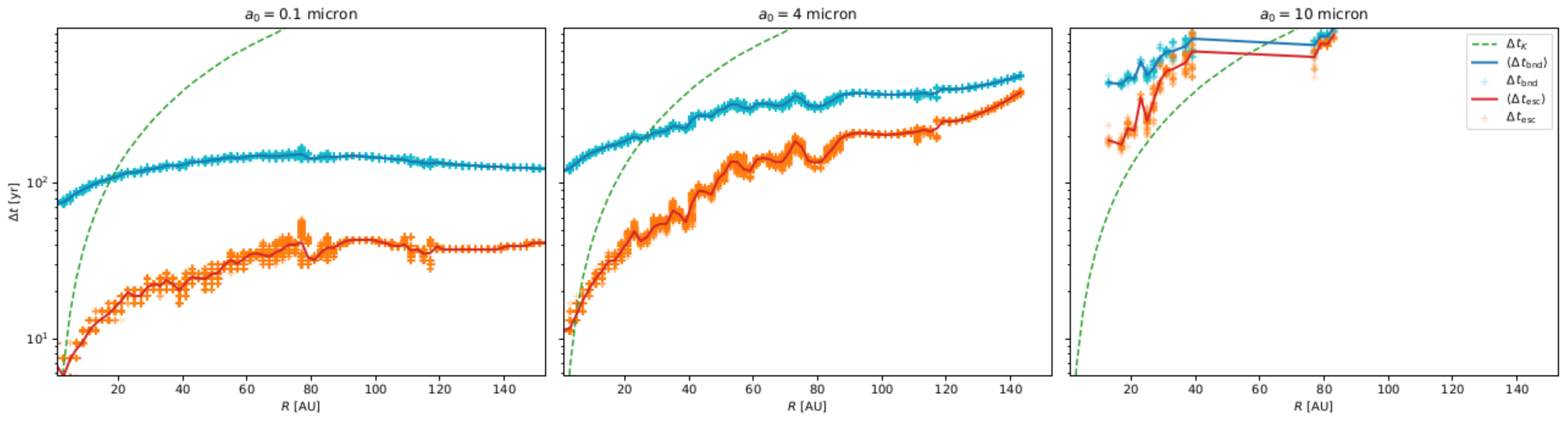}
    \caption{Time needed to fully blow out dust particles from when they first enter the wind at $t_\mathrm{wind,0}$ to the domain boundary, which they reach at $t_\mathrm{bnd}$ (data points in cyan, mean in blue), and time needed to accelerate dust grains to $v_\mathrm{r,esc}$ (data points in orange, mean in red); both for a selection of three $a_0$.
    Keplerian orbital times ($t_\mathrm{dyn} \equiv \Delta t_K$) at the disk surface are included as dashed green lines.
    The rasterization of the data points results from a time-discrete particle tracking and a binning in $R$-direction; thus, one raster point may represent multiple data points.
    (For all $a_0$, see Fig.~\ref{fig:timescales-vel-all}.)}
    \label{fig:timescales-vel-3}
\end{figure*}

In Fig.~\ref{fig:trajectories-analysis}, we have included a timescale $\Delta t_\mathrm{bnd}$ for the motion of the dust grains.
This timescale was computed as the difference between the time at which the particle is picked up by the wind ($t_\mathrm{wind,0}$) and the time at which it crosses the domain at $r > 300\,$AU while being entrained ($t_\mathrm{bnd}$), thus $\Delta t_\mathrm{bnd} = t_\mathrm{bnd} - t_{\mathrm{wind},0}$.
The full distribution of timescale data points for all trajectories is shown in Fig.~\ref{fig:timescales-vel-3} (cyan and blue).

In general, wind entrainment timescales appear to span $10^2$ to $10^3$\,yr.
This is similar to the range given by \citet{Birnstiel-2016} for midplane dust growth; since dust densities at $z \gtrsim 3.5\,H$ (see Fig.~\ref{fig:gasdisk}) are considerably lower, dust-dust interactions should indeed be negligible, as claimed in Section~\ref{sec:Methods}.
More concisely, \citet{Kornet-2001} have established that the dust growth timescale goes as $t_\mathrm{grow} \propto \varrho_\mathrm{dust} \cdot v_\mathrm{dust}$, with $\varrho_\mathrm{dust}$ the local dust density.\footnote{$\varrho_\mathrm{dust}$ is not to be confused with the internal grain density $\varrho_\mathrm{grain}$.}
As Table~\ref{tab:rho-v-gas} shows, $\varrho_\mathrm{gas}$ drops off heavily towards the disk surface while $v_\mathrm{gas} \equiv \left| \vec{v}_\mathrm{gas} \right|$ remains very comparable; Figs.~\ref{fig:gasdisk} and \ref{fig:dust-positions} illustrate that the wind speed picks up slightly above -- and not at -- the base of the flow (as would be expected considering the latter is located slightly below the disk surface).
Hence the dust-dust interaction timescale is much longer than the wind blowout time, if we assume a constant dust-to-gas ratio.
In case of a more realistic relation, dust would be even more scarce than gas for similar $z$ \citep{Krijt-2016}.

\begin{table}
    \centering
    \caption{Comparison of gas densities and velocities at the disk midplane and the base of the XEUV flow.}
    \label{tab:rho-v-gas}
    \begin{tabular}{lcrr}
        & & \multicolumn{1}{l}{10\,AU} & \multicolumn{1}{l}{100\,AU} \\ \hline
        $\varrho_\mathrm{gas} \,[\mathrm{g/cm^3}]$ & midplane & $8 \cdot 10^{-13}$ & $8 \cdot 10^{-16}$ \\
                                                   & flow base  & $2 \cdot 10^{-17}$ & $2 \cdot 10^{-19}$ \\
        $\left| \vec{v}_\mathrm{gas} \right| \,[\mathrm{km/s}]$ & midplane & 8 & 2 \\
                                                                & flow base  & 8 & 2
    \end{tabular}
\end{table}

The $0.1\,\mu$m grains of Fig.~\ref{fig:timescales-vel-3} have $\Delta t_\mathrm{bnd}$ as low as 70\,yrs if they are picked up at small $R$; this means that the longer distance to the domain boundary is outweighed by the higher acceleration the dust experiences close to the star.
For $R \gtrsim 80\,$AU, $\Delta t_\mathrm{bnd}$ decreases; if we assume the general trend of slower speed-up from larger $R$ to persist, this would mean that the grains are picked up close enough to the computational boundary (at $R \simeq 300\,$AU) to be blown out faster than those from slightly further in.
So this drop-off is caused by the numerical setup, not by the actual physics involved.

For $a_0 \geq 0.5\,\mu$m, we do not see this fall-off anymore.
Disregarding the various local peaks in $\Delta t_\mathrm{bnd}$ which are caused by the base of the wind not being perfectly smooth,\footnote{This can be seen from a close examination of the black lines in Fig.~\ref{fig:dust-positions}, as well as Fig.~\ref{fig:gasdisk-surface}.} a clear trend of the blow-out time $\Delta t$ increasing with the launching position $R$ emerges.
For $10\,\mu$m grains, we find $\max(\Delta t) \approx 10^3\,$yr, which is still well below the simulation time frame $\Delta t_\mathrm{sim}$, validating the latter a posteriori.

Since the cutoff at $r \simeq 300\,$AU is somewhat arbitrary, the orange and red parts of Fig.~\ref{fig:timescales-vel-3} show a different approach to defining a timescale:
These represent the time between wind pick-up ($t_\mathrm{wind,0}$, as above) and reaching $v_{r,\mathrm{esc}}$ at $t_\mathrm{esc}$, that is $\Delta t_\mathrm{esc} = t_\mathrm{esc} - t_{\mathrm{wind},0}$.
Because the velocity field of the gas flow in the wind is pointing outwards (see Fig.~\ref{fig:gasdisk}), it is highly unlikely that a grain will not be fully blown out by the XEUV wind once it has reached $v_{r,\mathrm{esc}}$ .

The overall appearance of the average values for $\langle \Delta t_\mathrm{bnd} \rangle$ (blue) and $\langle \Delta t_\mathrm{esc} \rangle$ (red) is quite similar; both show a rather distinct upwards trend, mitigated only for $R \gtrsim 80\,$AU and $a_0 < 0.5\,\mu$m.
Interestingly, this feature also holds for $\Delta t_\mathrm{esc}$, which in contrast to $\Delta t_\mathrm{bnd}$ does not depend on the choice of the computational boundary.

The values retrieved for $\Delta t_\mathrm{esc}$ are different -- and always smaller.
This difference is most pronounced for small grains, which reach $v_{r,\mathrm{esc}}$ within 5 to 60\,yr (for $a_0 = 0.1\,\mu$m), while needing 70 to 170\,yr to leave the simulation domain.
Dust particles with $a_0 \leq 8\,\mu$m launched close to the star are accelerated to $v_{r,\mathrm{esc}}$ within merely a few years; bigger grains take longer to pick up speed, or are too heavy to be picked up by the wind at all ($a_0 \geq 12\,\mu$m).

The dashed green lines in Fig.~\ref{fig:timescales-vel-3} indicate the steady-state dynamical timescale $t_\mathrm{dyn} = 2\,\pi / \Omega_K = 2\,\pi / \sqrt{G\,M_* / r^3}$ for a Keplerian orbit at the base of the XEUV-driven flow.
For $a_0 = 0.1\,\mu$m, $t_\mathrm{dyn}$ dominates the dust motion only for starting points $R \lesssim 20\,$AU; this increases to $R \lesssim 30\,$AU for $a_0 = 4\,\mu$m.
By contrast, the photoevaporation of the $10\,\mu$m grains is largely dominated by $t_\mathrm{dyn}$, which for those is mostly larger than $\Delta t_\mathrm{bnd}$.

Because $\Delta t_\mathrm{esc} < \Delta t_\mathrm{bnd}$, we find $t_\mathrm{dyn} > \Delta t_\mathrm{esc}$ for both $0.1\,\mu$m and $4\,\mu$m for $R > 10\,$AU, meaning that for almost all grains the blow-out happens (considerably) faster than their `usual' timescale.
It is only for the large particles ($a_0 = 10\,\mu$m) that the time needed for acceleration to the escape velocity becomes comparable to the Keplerian timescale.

\subsection{Maximum entrained grain size}

\begin{figure}
    \centering
    \includegraphics[scale=0.625]{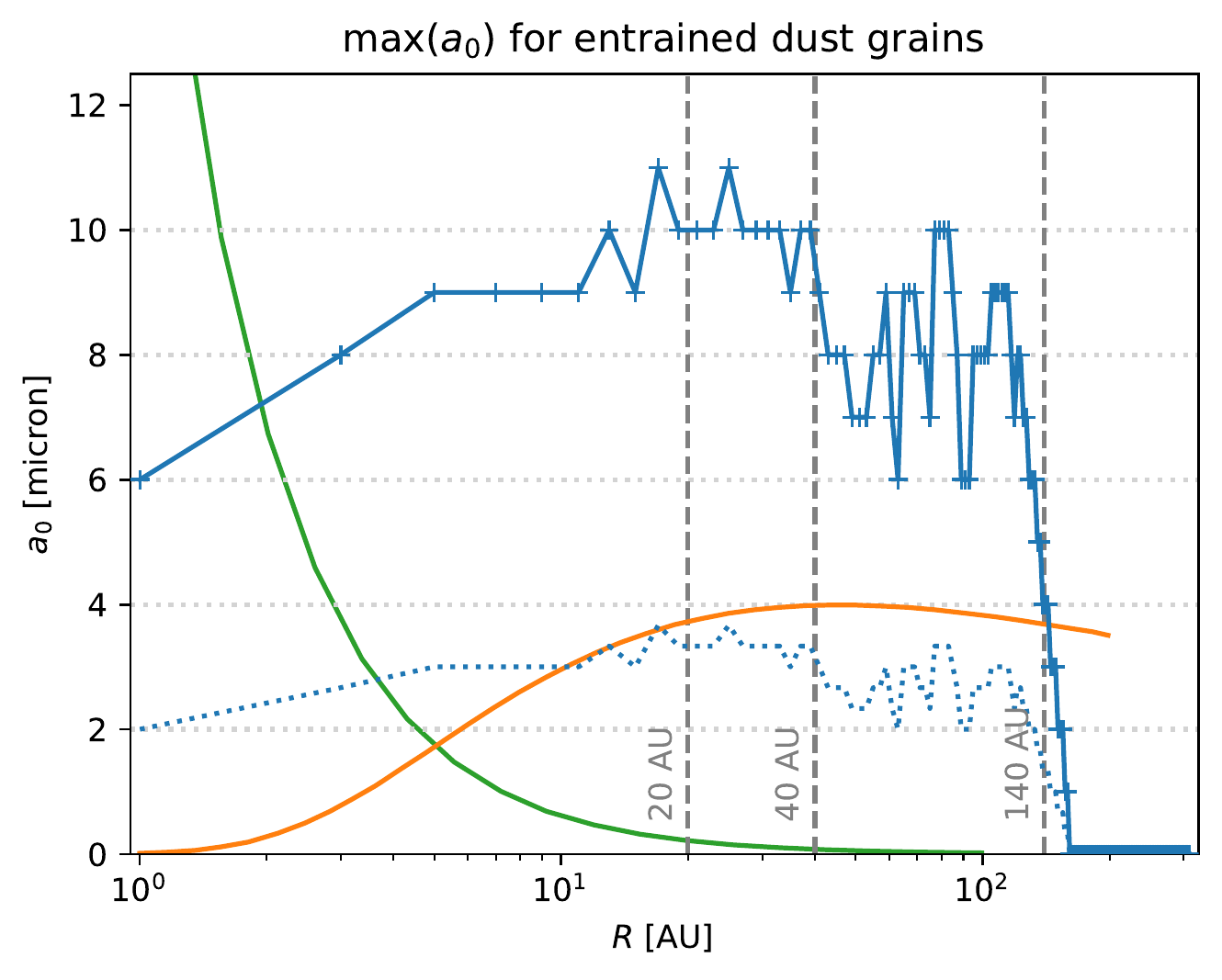}
    \caption{Size of the largest grains entrained from a point $R$ along the base of the wind (blue, peak at about 20\,AU); the saw-tooth appearance of the curve at larger $R$ is caused by the finite resolution of the underlying gas grid.
    When comparing to \citet[][their Fig.~7 for $M_*=0.75\,\mathrm{M}_\odot$]{Hutchison-2016c} (orange, peak at around 40\,AU), we can see the size enhancement -- especially at smaller $R$ -- caused by the inclusion of X-rays in our photoevaporative wind model.
    The blue dashed line represents our results scaled down by a factor of 3, to compensate for the differing internal grain densities of \citet{Hutchison-2016c} and this work; yet, $\dot{M}_\mathrm{w}$ still differs between the models, making a direct comparison difficult.
    The MHD wind model investigated by \citet[][their Fig.~4]{Miyake-2016} (green) shows a distinctly different entrainment curve, starting off at very high $a_0$ in the jet region but dropping towards $\max(a_0) = 0$ very quickly.
    Around $R \simeq 140\,$AU, our $\max(a_0)$ plummets to 0.}
    \label{fig:max(a0)-1D}
\end{figure}

As has already been suggested in Fig.~\ref{fig:dust-positions}, dust grains may be too heavy to be blown out by the XEUV wind.
This is investigated further in Fig.~\ref{fig:max(a0)-1D} where the blue line shows the maximum grain size $\left. \max(a_0) \right|_{R}$ that can be fully entrained in the wind from a starting position $R$ along the base of the flow.
As noted in Section~\ref{sec:Methods}, the latter is not entirely smooth; this, in turn, causes the craggy appearance of the graph for $\max(a_0)$.

To facilitate direct comparisons, we have included the corresponding EUV-only curve of \citet{Hutchison-2016c} in orange and the MHD-wind results of \citet{Miyake-2016} in green.
The figure shows that overall, an X-ray driven wind is able to entrain larger grains over a larger radial range than its EUV-only or MHD-driven counterparts; this may have a noticeable impact on deduced opacity maps, and also on the detectability of the wind in scattered light.

The biggest grains that are blown out, that is $a_0 = 11\,\mu$m, are entrained from $15 \lesssim R\,[\mathrm{AU}] \lesssim 30$.
Closer to the star, stellar gravity counteracts the gas drag force; but since the gravitational pull drops off with $r^2$, the drag force dominates particle motion further out.\footnote{This correlates to \citet{Clarke-2016} limiting the applicability of their scale-free gas motion to $R \gg R_g$, with $R_g$ the gravitational radius.}
With increasing $R$, $\max(a_0)$ slowly decreases out to $R_\mathrm{max} \simeq 140\,$AU, where it quickly drops to $\max(a_0)=0$.
This coincides with the maximum radius at which XEUV photoevaporation is effective for the gas component of the disk; \citet{Picogna-2019} show that the surface mass-loss rate ($\dot{\Sigma}_\mathrm{gas}$) drops to negligible values at $R \approx 140\,$AU.
So both gas and dust residing at the disk surface at $R \gtrsim 140\,$AU are very unlikely to be thermally unbound from there.
Thus, for both their and our simulations this marks the outer boundary of the XEUV-wind-dominated region of the protoplanetary disk.
Furthermore, it validates a posteriori our choice to limit the computational domain to $r \lesssim 300\,$AU and the initial particle placement to $r \leq 200\,$AU.\footnote{The base of the flow is at $z \approx 103\,$AU at $R = 140\,$AU (see Fig.~\ref{fig:gasdisk}), yielding $r \approx 174\,\mathrm{AU} < 200\,$AU.}

As mentioned in Section~\ref{sec:Methods}, $\varrho_\mathrm{grain}$ is not well-constrained.
A variation of the dust density is found to directly anti-correlate with $\max(a_0)$, that is $\varrho_\mathrm{grain} \propto 1/\max(a_0)$; this agrees with the analytical findings of \citet[][their Eq.~(15)]{Hutchison-2016c}.
Hence, for a threefold internal grain density of $\varrho_\mathrm{grain}' = 3\,\mathrm{g/cm}^3 = 3\,\varrho_\mathrm{grain}$, our global $\max(a_0)$ drops to $\max(a_0)' = \max(a_0) / 3 \approx 3.5\,\mu$m.
An accordingly scaled version of our results is included in Fig.~\ref{fig:max(a0)-1D} as the blue dashed line in order to allow for simpler comparison to the results of \citet{Hutchison-2016c, Hutchison-2016b}; but it should be kept in mind that their and our mass-loss rates are not entirely similar.
While they quote a surface mass-loss rate of $\dot{\Sigma}_\mathrm{gas} = 3.6 \cdot 10^{-12}\,\mathrm{g/(cm^2\,s)}$ at $R=5\,$AU, we have $\dot{\Sigma}_\mathrm{gas} \approx 2 \cdot 10^{-13}\,\mathrm{g/(cm^2\,s)}$ \citep[][their Fig.~5]{Picogna-2019}.

\begin{figure}
    \centering
    \includegraphics[scale=0.55]{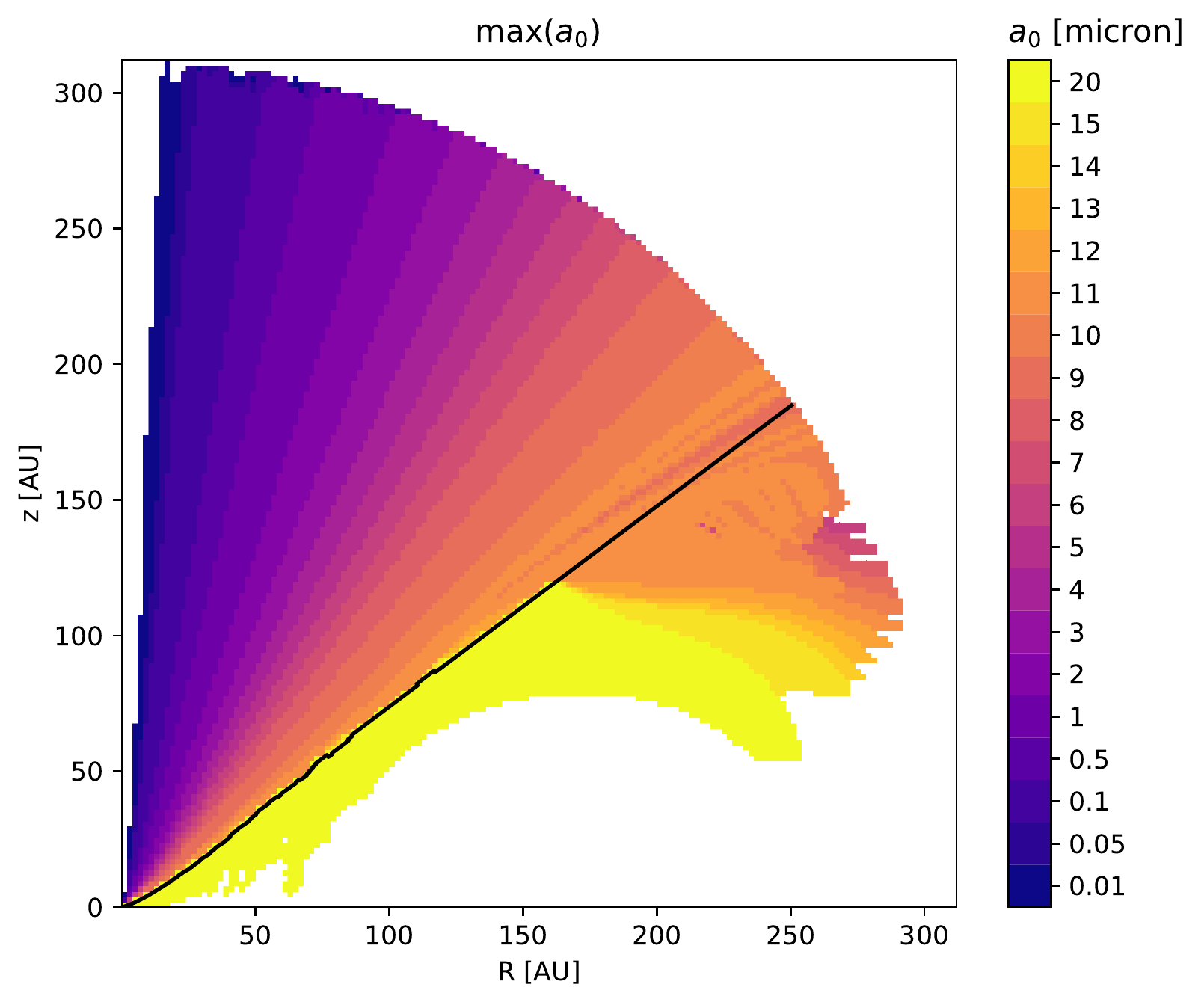}
    \caption{Maximum size $\max(a_0)$ of the dust grains in the wind in analogy to \citet[][their Fig.~2]{Owen-2011a}.
    The brightest color found above the sonic surface indicates a global maximum of $\max(a_0) = 11\,\mu$m; the region below the base of the wind (black) is included merely for completeness.
    The visible correlation between grain size and maximum scale height is further investigated in Fig.~\ref{fig:N(R,z)-3}.}
    \label{fig:max(a0)-2D}
\end{figure}

A 2D map of the grain sizes that can populate different regions of the wind is shown in Fig.~\ref{fig:max(a0)-2D}.\footnote{For this, the particle motions were mapped to a 2\,AU\,$\times$\,2\,AU grid; a much higher resolution would introduce artifacts due to insufficient particle count whereas a lower one would smear out features.}
The larger grains remain rather close to the disk surface; by contrast, smaller ones are lifted up to larger scale heights.
For the same launching position $R$, smaller grains reach higher $z$.
The abrupt decline of the $20\,\mu$m grains at $R \gtrsim 160\,$AU results from the initial placement of the particles within $0.33 \leq r\,[\mathrm{AU}] \leq 200$, or equivalently $0.3 \lesssim R\,[\mathrm{AU}] \lesssim 160$.

In Fig.~\ref{fig:N(R,z)-3}, we see the regions which are populated by grains of a given $a_0$ in green. We find that there are no deserts of smaller dust particles in regions populated by larger ones; for instance, wherever we find grains with $a_0 = 10\,\mu$m, we also find grains with $a_0 = 0.1\,\mu$m and $a_0 = 4\,\mu$m.
From Fig.~\ref{fig:max(a0)-2D}, we have learnt that smaller grains will reach higher scale heights; to quantify this behaviour, we have included the maximum height $\left. \max(z) \right|_R$ for a certain $a_0$ at $R$ in orange in Fig.~\ref{fig:N(R,z)-3}.
Fits with a simple second-order polynomial,
\begin{equation}\label{eq:fit-max(z)}
    \left. \max(z) \right|_R = c_1 \, R + c_2 \, R^2 \;,
\end{equation}
are shown as blue dotted lines; they match $\left. \max(z) \right|_R$ quite well except for the very inner region where the dust distribution is slightly more flared.\footnote{Note that omitting disk gravity should not have a strong effect on the strength of the flaring seen here, see Appendix~\ref{sec:app:disk-gravity}.}
Higher-order polynomials fit the inner region better, however they are not included here since these fits are not a physically-derived, but merely a numerical prescription which we intended to keep quite simple.

\begin{figure*}
	\centering
	\includegraphics[scale=0.4]{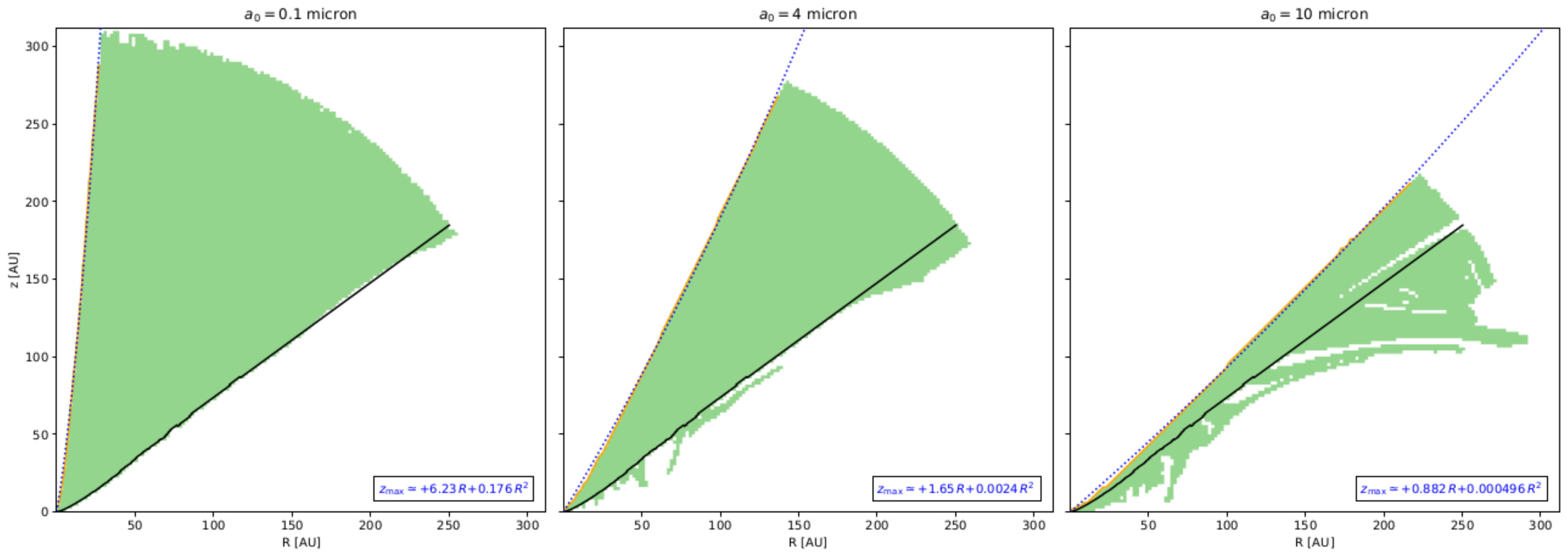}
	\caption{Areas populated by the dust grains (green); since we do not start from a realistic distribution along the base of the wind, we do not portray a density map.
	Smaller grains reach higher $z$ at similar $R$.
	Wind base in black, numerical $\left. \max(z) \right|_R$ in orange, and corresponding fit in (dotted) blue; fits according to Eq.~\eqref{eq:fit-max(z)} annotated.
	(For all $a_0$, see Fig.~\ref{fig:N(R,z)-all}.)}
	\label{fig:N(R,z)-3}
\end{figure*}

The fit parameters $c_1$ and $c_2$ for all grain sizes are shown in Fig.~\ref{fig:N(R,z)-fitparam} together with merely phenomenological prescriptions for the scaling of $c_1$ and $c_2$ with $a_0$ which are intended for comparisons to observational data from edge-on disks in future work.
The fit formulas used are meant solely for a simplified reproduction of the values of $c_1$ and $c_2$; they are not based on physical considerations.

\begin{figure}
	\centering
	\includegraphics[scale=0.5]{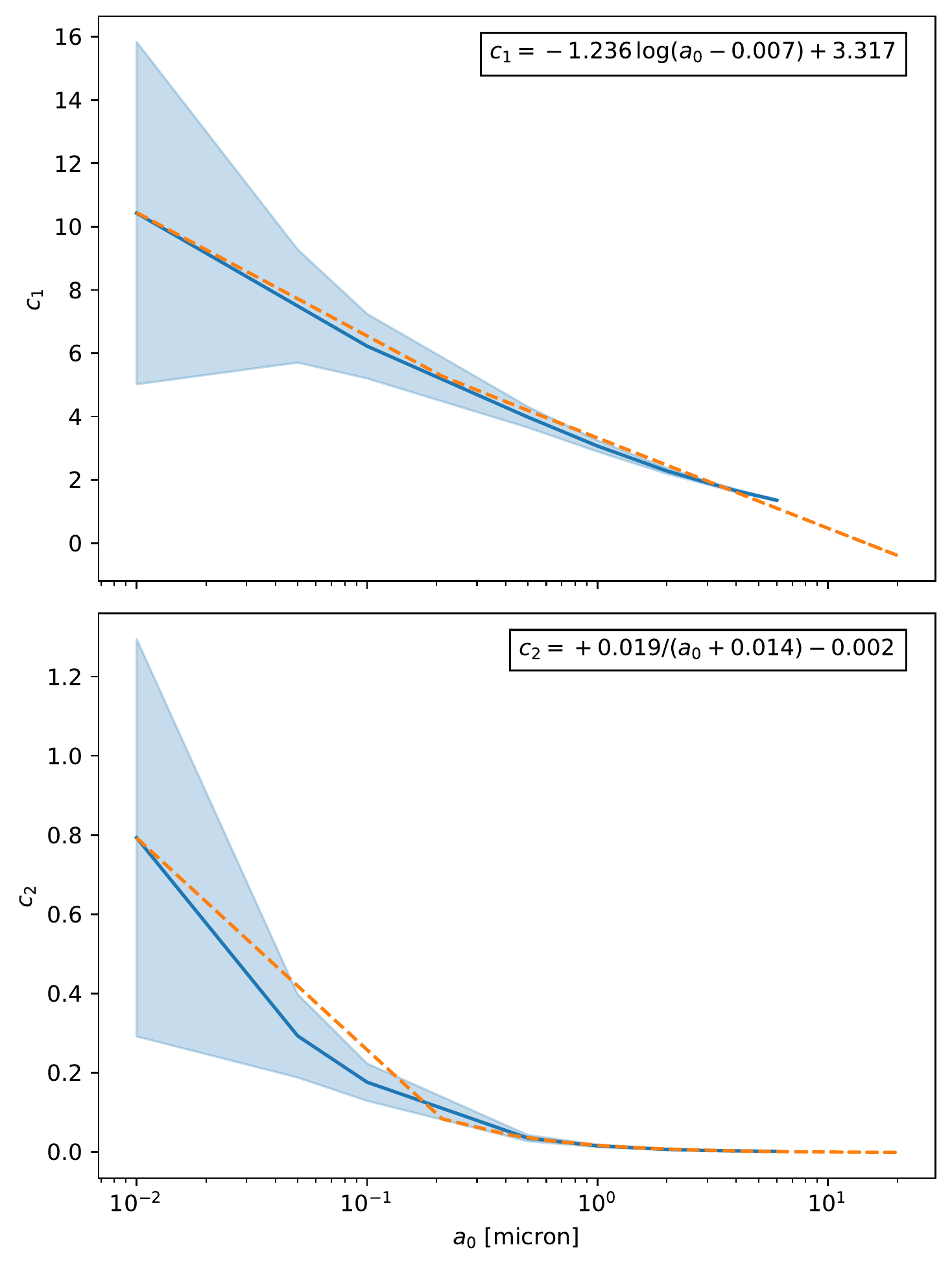}
	\caption{Fit parameters for all fits for $\left. \max(a_0) \right|_R$ as a function of $a_0$ in blue, with $3\,\sigma$-errors included as shaded regions. Non-physical fits to these fit parameter curves in orange, with the parameterizations given in the text boxes.}
	\label{fig:N(R,z)-fitparam}
\end{figure}

As may have been expected, both $c_1$ and $c_2$ decrease with increasing $a_0$, indicating the decline of the slope of $\left. \max(z) \right|_R$ with $a_0$ already seen in Fig.~\ref{fig:N(R,z)-3}.
The higher errors at lower $a_0$ result from the larger inclination of $\left. \max(z) \right|_R$ for smaller grains; in addition, the population map shown in Fig.~\ref{fig:N(R,z)-3} has a finite resolution as outlined above, which further contributes to a larger uncertainty especially for very steep lines.

Furthermore, we find $c_1 \gg c_2$ for all $a_0$ modelled; this corresponds to the mostly linear appearance of $\left. \max(z) \right|_R$ especially for larger $R$.
Yet $c_2$ does not drop to zero, so a certain amount of flaring is preserved for all particle sizes.


\section{Discussion}
\label{sec:Discussion}

We have numerically simulated the trajectories of dust grains in the XEUV-irradiated wind regions of a gaseous protoplanetary disk.
As was to be expected \citep[see e.g.][]{Armitage-2015}, we found small dust particles to show very good agreement with the gas streamlines and, in contrast, bigger grains to noticeably deviate from them. Thus, analytical models of gas motion as provided by \citet{Clarke-2016} cannot be used to accurately model the trajectories of dust grains attaining $St \gtrsim 0.5$ in the wind.

\subsection{Photoevaporative winds and radiation pressure}

\citet{Owen-2019} use the X-ray photoevaporation model of \citet{Owen-2011b} to quantify grain blow-out due to direct radiation pressure alone; since their $M_*$, $M_\mathrm{disk}$, and $L_X$ are the same as used in this work and their internal grain density of $1.25\,$g/cm$^3$ only slightly differs from our $\varrho_\mathrm{grain} = 1\,\mathrm{g/cm^3}$, this allows for an almost perfect comparison of their results to grain entrainment by XEUV winds.\footnote{\citet{Picogna-2019} provide a more in-depth explanation of the differences between their model and that of \citet{Owen-2011b}.}
Using an effective stellar surface temperature of 4500\,K, they find $\max_\mathrm{rad}(a_0) \simeq 0.6\,\mu$m for the largest grains for which the radiation pressure from an equivalent black body still outweighs stellar gravity.
Thus, a photoevaporative XEUV wind enhances the size of dust grains blown out by a factor of almost 20 over radiation pressure alone.

Hence we can conclude that a photoevaporative XEUV-driven wind is much more effective at removing larger dust grains than direct radiation pressure.

\subsection{Comparison to MHD wind models}

\citet{Miyake-2016}, building on an MHD model established by \citet{Suzuki-2009}, performed 1D simulations of dust motion in MHD-driven winds.
While they mainly focused on floating grains,\footnote{By `floating grains', \citet{Miyake-2016} refer to dust particles of sizes $25 \lesssim a_0\,[\mu\mathrm{m}] \lesssim 45$ which float near the sonic surface of the disk, neither too heavy to fall back down towards the midplane nor light enough to be blown out by their MHD wind.} they also give a maximum entrainable grain size for their model along $R$; we have included it in Fig.~\ref{fig:max(a0)-1D} in orange.
This shows that whereas MHD winds excel at removing large dust grains from regions very close to the star, photoevaporative winds start to dominate -- in terms of entrained grain size -- at $R \gtrsim 2\,$AU.
Thus, MHD winds would seem to be limited to the jet region and its immediate surroundings.

Recently, \citet{Giacalone-2019} have investigated dust transport in (cold) magneto-centrifugally-driven disk winds in 2D; they conclude that the region of interest for dust pick-up from and re-deposition on the disk surface covers their full modelled range of $R$, which they chose to set up as $0.1 \leq R \,[\mathrm{AU}] \leq 100$.
For their T-Tauri model, they find entrainment of grains with $a_0 \lesssim 2\,\mu$m (their Fig.~3).
This value was obtained for a disk surface temperature of $T_\mathrm{surf} = 600\,$K, which seems very plausible from an MHD point of view, but somewhat low from a photoevaporative one.
They show as well that $\max(a_0)$ clearly depends on said $T_\mathrm{surf}$, with higher $T_\mathrm{surf}$ increasing their $\max(a_0)$.

\begin{figure}
	\centering
	\includegraphics[scale=0.5]{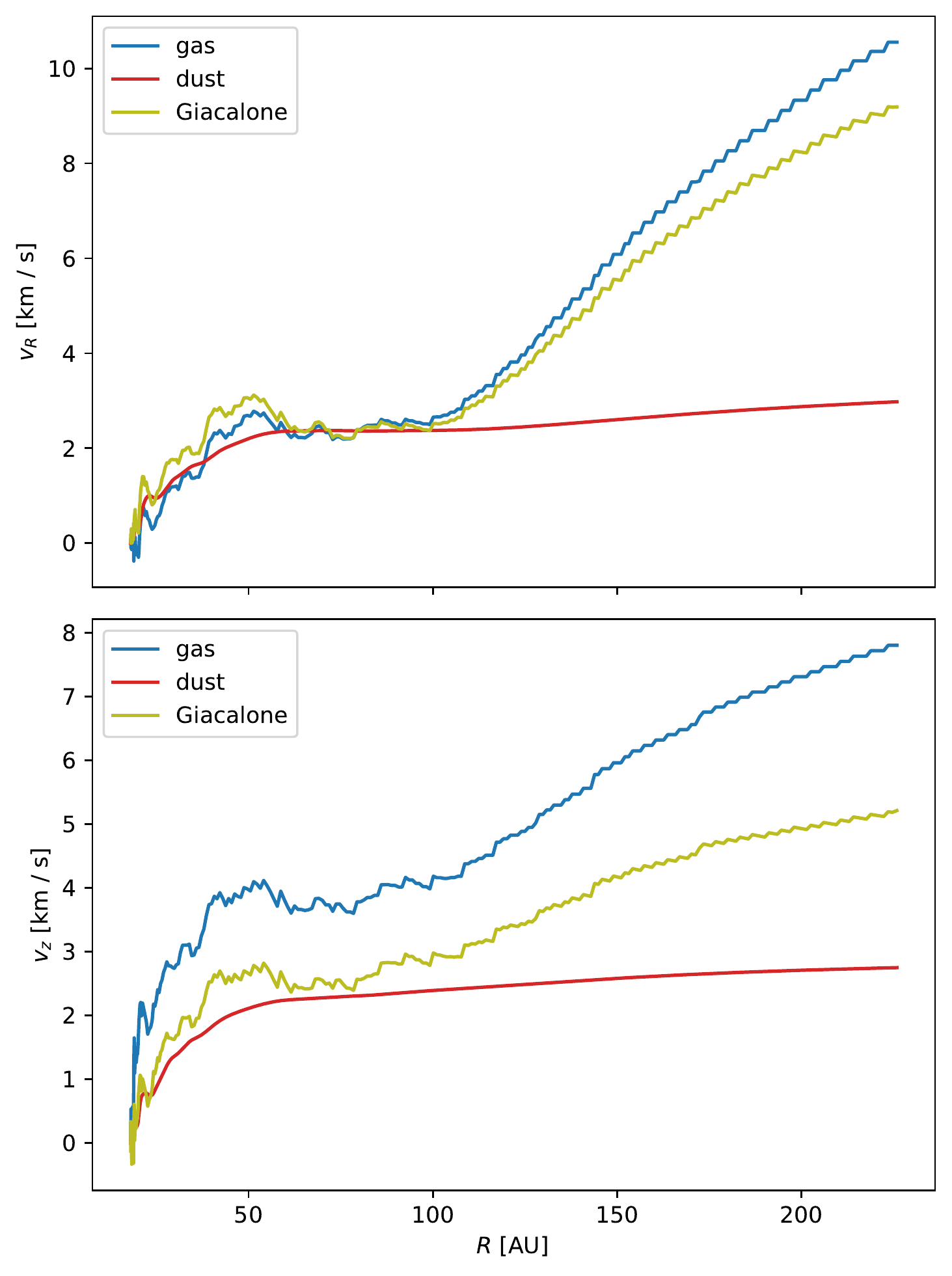}
	\caption{Re-plot of $v_R$ and $v_z$ for the same dust grain as in the right column of Fig.~\ref{fig:trajectories-analysis} ($a_0 = 10\,\mu$m).
	The dust parameters are colored red, the surrounding gas in blue; the motion according to Eqs.~(1) and (3) of \citet{Giacalone-2019}, with $t_\mathrm{stop}$ as retrieved from our model, is added in green.
	Especially towards larger $R$, the semi-analytical prescription does not reproduce the dust motion very well.
	It should be noted, however, that \citet{Giacalone-2019} do not use it for grains this big.}
	\label{fig:trajectories-Giacalone}
\end{figure}

In contrast to our numerical setup, \citet{Giacalone-2019} opted for a semi-analytical approach to trajectory modelling (see their Eqs.~(1) through (3)); Fig.~\ref{fig:trajectories-Giacalone} compares the grain velocities we extract for the $10\,\mu$m grain of Fig.~\ref{fig:trajectories-analysis} to their prescription.
We are only showing a comparison for a large dust particle here because for small $a_0$, gas and dust velocities are very similar.
It should be noted, in this regard, that \citet{Giacalone-2019} use their semi-analytical equations just for relatively small dust grains, considering their $\max(a_0)$ as quoted above; we present Fig.~\ref{fig:trajectories-Giacalone} merely to show that numerical simulations are still necessary for modelling the trajectories of the larger grains whose velocity distinctly decouples from the gas flow.
Thus, until more intricate analytical prescriptions are introduced for tracing dust motion in a photoevaporative wind, a numerical approach must be employed for accurate results.

\subsection{Comparison to EUV wind models}

Previous investigations of dust entrainment in photoevaporative winds have limited themselves to EUV-only photoevaporation \citep{Owen-2011a, Hutchison-2016c}.

\citet{Owen-2011a} base their work on an EUV-luminosity optimized model of a $2.5\,\mathrm{M}_\odot$ Herbig Ae/Be star; while this means that our results are not directly comparable in terms of stellar parameters, both their and our work test the respective highest-luminosity (i.e. best-case) scenario, and hence aim to provide an upper limit to $\max(a_0)$.
Comparing their $\max(a_0) \simeq 2.2\,\mu$m (see their Fig.~2) to our value of $11\,\mu$m, both retrieved for $\varrho_\mathrm{grain} = 1\,\mathrm{g/cm^3}$, we find a clear enhancement of particle sizes blown out by the wind.
Thus, depending on the mass-loss rates caused by XEUV photoevaporation, the inclusion of X-ray photons in the disk irradiation model may be a crucial component for accurately predicting dust entrainment.

The basic mechanics do not change, though; when comparing our Fig.~\ref{fig:max(a0)-2D} to \citet[][their Fig.~2]{Owen-2011a}, we see the same qualitative behaviour.
In both cases, the maximum height which a grain of a given $a_0$ at a given $R$ can be lifted to decreases with increasing $a_0$. As expected, the largest entrainable grains are not lifted up very high above the disk by the photoevaporative wind.

\citet{Hutchison-2016c} explore a wide range of stellar parameters; apart from choosing a penetration depth typical for T-Tauri stars \citep[provided by][]{Woitke-2016}, they also ran their models for a stellar mass of $0.75 \, \mathrm{M}_\odot$, very similar to our $M_* = 0.7\,\mathrm{M}_\odot$.
Just like us, they find an anti-correlation $M_* \propto 1/\max(a_0)$ which explains their value of $\max(a_0) \simeq 4\,\mu$m in comparison to \citet{Owen-2011a}.

This value is still distinctly smaller than the $11\,\mu$m we have found to be picked up; yet as already noted in Section~\ref{sec:Results}, we get very close to their EUV-only results with our model when using their $\varrho_\mathrm{grain} = 3\,\mathrm{g/cm^3}$.
However, as pointed out above as well, the entrainable grain sizes are still not comparable since gas mass-loss rates strongly differ between \citet{Hutchison-2016c, Hutchison-2016b} and our model.
This is due merely to the different numerical setup of their and our models; \citet{Owen-2012a} and \citet{Owen-2012c} show that for viable $L_X$, X-ray driven winds clearly dominate over EUV-driven ones in terms of $\dot{M}_\mathrm{w}$.
In addition to this, we have now seen that even at lower $\dot{\Sigma}_\mathrm{gas}$, XEUV winds may entrain larger grains than EUV-only ones.

Furthermore, Fig.~\ref{fig:max(a0)-1D} shows that independent of the exact grain size, X-ray irradiation shifts the peak in $\max(a_0)$ towards lower $R$.
The T-Tauri star of \citet{Hutchison-2016c} entrains its $\max(a_0)$ from $40 \lesssim R\,[\mathrm{AU}] \lesssim 50$, and the more luminous Herbig Ae/Be star of \citet{Owen-2011a} from $30 \lesssim R\,[\mathrm{AU}] \lesssim 40$.
Interestingly, albeit exerting a stronger gravitational pull, the higher-mass star blows out its largest grains from further in; thus it seems that the higher gravity is outweighed by an even stronger wind launched.

In comparison, our XEUV wind picks up its largest grains from $20 \lesssim R\,[\mathrm{AU}] \lesssim 30$, indicating that dust grains are entrained more efficiently closer to the star.
In contrast to \citet{Giacalone-2019} and their MHD wind model, we only find grain fallback for $R > 200\,$AU ($a_0 = 0.1\,\mu$m), or $R \gtrsim 150\,$AU ($a_0 = 10\,\mu$m; see Fig.~\ref{fig:N(R,z)-3}) for launching positions within a similar range of $R$ -- as noted above, the trajectories do not intersect (see Fig.~\ref{fig:trajectories-samples-3}). Hence, the photoevaporative wind re-deposits only little material on the disk surface.
This agrees with the conclusions of \citet{Owen-2011a} that once entrained, a dust grain will almost always remain in the wind, and be carried out to large radii.
Furthermore, as noted above, all grains leaving the computational domain above the disk surface have $v_r > v_{r,\mathrm{esc}}$, and are hence very likely to entirely leave the protostellar environment.


\section{Summary}
\label{sec:Summary}

We have modelled dust trajectories for grain sizes $10^{-3} \leq a_0\,[\mu\mathrm{m}] \leq 10^2$ in the wind region of a $M_\mathrm{disk} \simeq 10^{-2} \,M_*$ gas disk around a $M_* = 0.7\,\mathrm{M}_\odot$ T-Tauri star, irradiating its surroundings with $L_X = 2 \cdot 10^{30}$\,erg/s on top of an EUV spectrum.
Our main findings are as follows:

\begin{itemize}
    \item X-ray driven winds are able to entrain grains up to a size of $a_0 \lesssim 11\,\mu$m; this is larger than the maximum entrained grain size from EUV-only models.
    
    \item XEUV winds pick up the largest particles from $R \simeq 20\,$AU.
    By contrast, EUV-only winds entrain their largest dust from further out (i.e. $R \simeq 40\,$AU). MHD winds show a very different profile, picking up very large grains from regions very close to the star ($R \ll 3\,$AU), but then rapidly loosing momentum farther from the star.
    
    \item Dust grains are launched with Stokes numbers $St < 0.4$ (i.e. $St \ll 1$).
    Once entrained, the large grains decouple from the gas flow; smaller dust particles decouple at later times.
    
    \item $\mu$m-sized dust grains of are blown out of the inner 300\,AU of a protoplanetary disk on a timescale of $10^2$ to $10^3$\,yr.
    
    \item For a given grain size, the launching point of a grain determines its further trajectory.
    
    \item Smaller dust grains may be lifted up higher by the wind, with the maximum height $\max(z)$ at a given $R$ decreasing with grain size $a_0$.
    
    \item An anti-correlation between $\left. \max(z) \right|_R$ and $a_0$ may be a typical signature of dusty photoevaporative winds.
\end{itemize}

Much like \citet{Owen-2011a}, we have found a strong dependence of the $\max(a_0)$ in the wind on $R$ (see Figs.~\ref{fig:max(a0)-2D} and \ref{fig:N(R,z)-3}).
This signature structure may be detectable in observations of edge-on disks if vertical mixing is strong enough to transport large grains to the disk surface; however, as has been shown by \citet{Hutchison-2016c}, this may not be the case, or may need additional MHD effects \citep{Miyake-2016}.

In a  future work we aim to present dust opacity maps and synthetic observations of typical protoplanetary dusty XEUV winds in order to investigate their detectability with current and future instrumentation.
As the mass-loss profiles of winds due to the different mechanisms -- that is, EUV, X-ray, MHD, etc. -- are shown (or expected) to be different, this will reflect in the size distribution and density of dust particles entrained at different locations in the wind.
While a quantitative discussion requires a calculation of detailed emission maps via radiative transfer modelling, we can speculate from the wind profiles of the gas that an X-ray-driven wind might produce a more extended launching region for larger grains than an EUV-only wind for which most of the entrainment is expected to occur near the gravitational radius of the disk.
The dust emission is expected to be concentrated even closer to the star in the case of an MHD wind.
Whether these differences are detectable with current instrumentation remains a matter of future investigation.


\begin{acknowledgements}

We would like to thank Cathie Clarke and Mark Hutchison for very helpful discussions, and the (anonymous) referee for a constructive report that improved the manuscript.
\\
This research was supported by the German Research Foundation (DFG, Deutsche Forschungsgemeinschaft), grants FOR 2634/1, ER 685/8-1, and ER 685/9-1, the Munich Institute for Astro- and Particle Physics (MIAPP) of the DFG cluster of excellence \textit{Origin and Structure of the Universe}.
T.B. acknowledges funding from the European Research Council (ERC) under the European Union's Horizon 2020 research and innovation programme under grant agreement No 714769.
The simulations have been carried out on the computing facilities of the Computational Center for Particle- and Astrophysics (C2PAP).

\end{acknowledgements}


\bibliographystyle{aa}
\bibliography{Draft.bib}


\begin{appendix}

\section{Omitting disk gravity}
\label{sec:app:disk-gravity}

As noted in Section~\ref{sec:Methods}, we decided not to include the gravitational pull from the gas disk in our model in order to cut computational costs. We verified a posteriori that this does not strongly impact our results.

To this end we compared, at each recorded position of all of our dust particles, the gas drag force
\begin{equation}\label{eq:Fdrag}
    F_\mathrm{drag,z} = \frac{\Delta v_\mathrm{gas,z} \, m}{t_\mathrm{stop}} \;,
\end{equation}
with a grain mass of $m = \frac{4\,\pi}{3}\,a_0^3\,\varrho_\mathrm{grain}$, to an overestimate of the gravitational pull the disk would produce,
\begin{equation}\label{eq:Fgrav}
    F_\mathrm{disk,z} = - \frac{G\,M_\mathrm{disk}\,m}{z^2} \;.
\end{equation}

\begin{figure}
    \centering
    \includegraphics[scale=0.60]{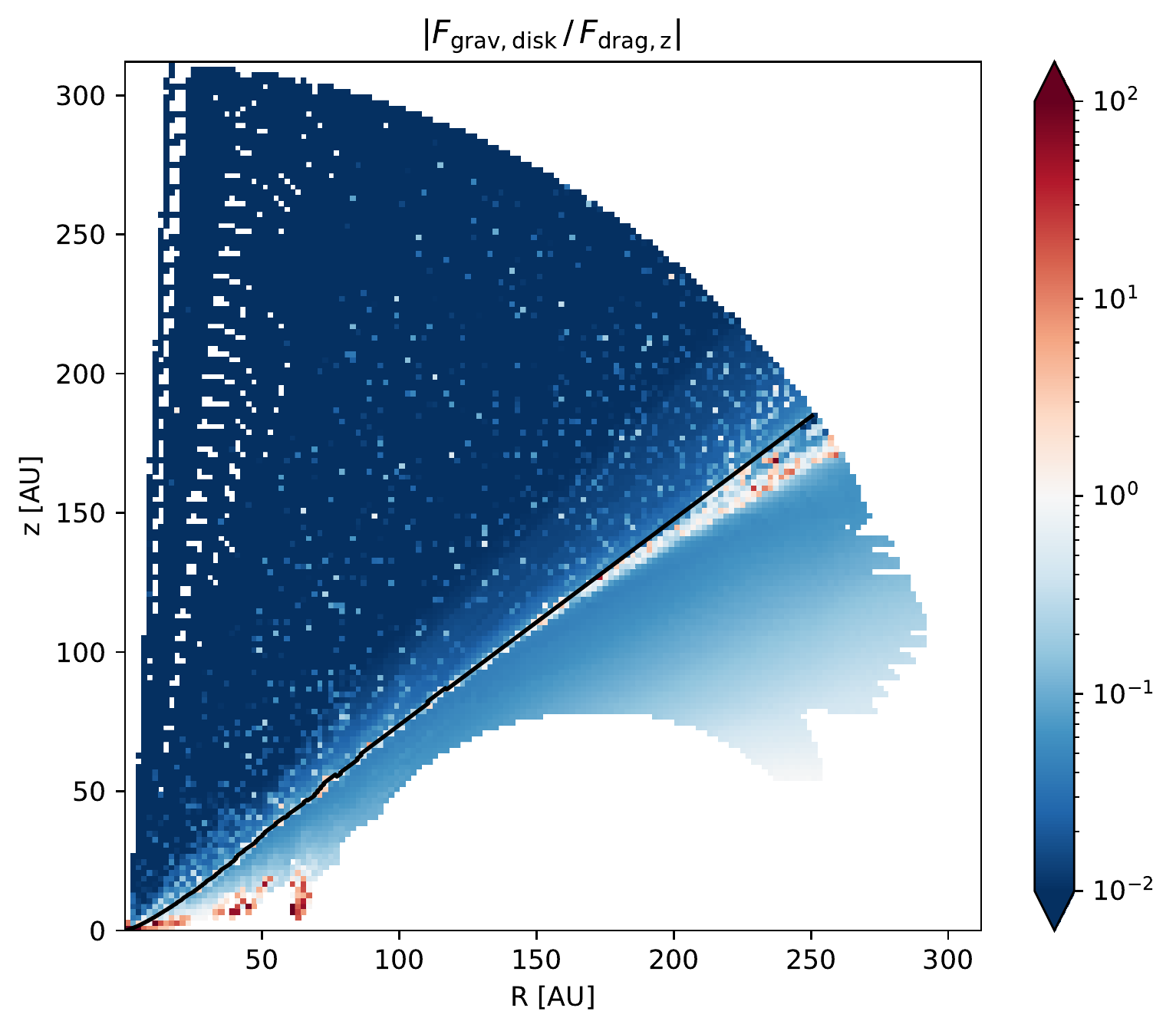}
    \caption{Maximum of the per-$a_0$ mean values for $|F_\mathrm{disk,z} / F_\mathrm{drag,z}|$ (see colorbar), mapped to a 2\,AU$\,\times\,$2\,AU grid. The gas drag clearly dominates over the (overestimated) disk gravity, especially in the wind region (base of the wind in black).
    Empty (white) regions above the sonic surface indicate that no particle has been recorded while in this cell, owing to the high grain speeds there (in contrast to Section~\ref{sec:Results}, we did not interpolate between recorded particle positions).}
    \label{fig:force-ratio-z}
\end{figure}

The latter formula leads to a slight overestimate of the $z$-component of $F_\mathrm{disk}$ because the disk is not centered exactly below a given particle.
If $F_\mathrm{disk,z} \ll F_\mathrm{drag,z}$, or rather $F_\mathrm{disk,z} / F_\mathrm{drag,z} \ll 1$, neglecting disk gravity should not noticeably affect our results.
To determine whether this is the case, we compute the mean value $\left\langle \left| F_\mathrm{disk,z} \,/\, F_\mathrm{drag,z} \right| \right\rangle$ of all grains of size $a_0$ on a 2\,AU$\,\times\,$2\,AU grid.
The maximum values (over all $a_0$) for all these means are shown in Fig.~\ref{fig:force-ratio-z}.
As the colors demonstrate, the gas drag clearly dominates over the overestimated gravitational pull from the protoplanetary disk.
We find only a few exceptions:
Firstly, well within the disk, from where it would be rather unlikely to see wind entrainment (as noted in Section~\ref{sec:Methods}, we have concentrated on realistically modelling the wind region, not the disk interior); secondly, very close to the host star, that is at $R \ll 10\,$AU, from where we do not see substantial entrainment (see Fig.~\ref{fig:max(a0)-1D}); and thirdly, around the base of the flow for $R \gtrsim 180\,$AU, which lies beyond the region from which wind pick-up happens in the first place, as noted in Section~\ref{sec:Results}.

In other words, we find that the photoevaporative wind is -- at low $r$ -- strong enough to compensate for both the closeness of the grain launching area to the disk midplane (and hence the center of mass) and also the concentration of the disk mass within $r \lesssim 100\,$AU (see Fig.~\ref{fig:gasdisk}).

At high $r$, the wind will not lose momentum (see the green velocity maps in Fig.~\ref{fig:dust-positions}), but the center of mass of the disk will have moved farther away from the base of the photoevaporative flow, meaning that gravity is even less likely to play a major role for wind entrainment.


\section{Including a spread in the initial velocities}
\label{sec:app:vel-sigma}

\begin{figure}
    \centering
    \includegraphics[scale=0.60]{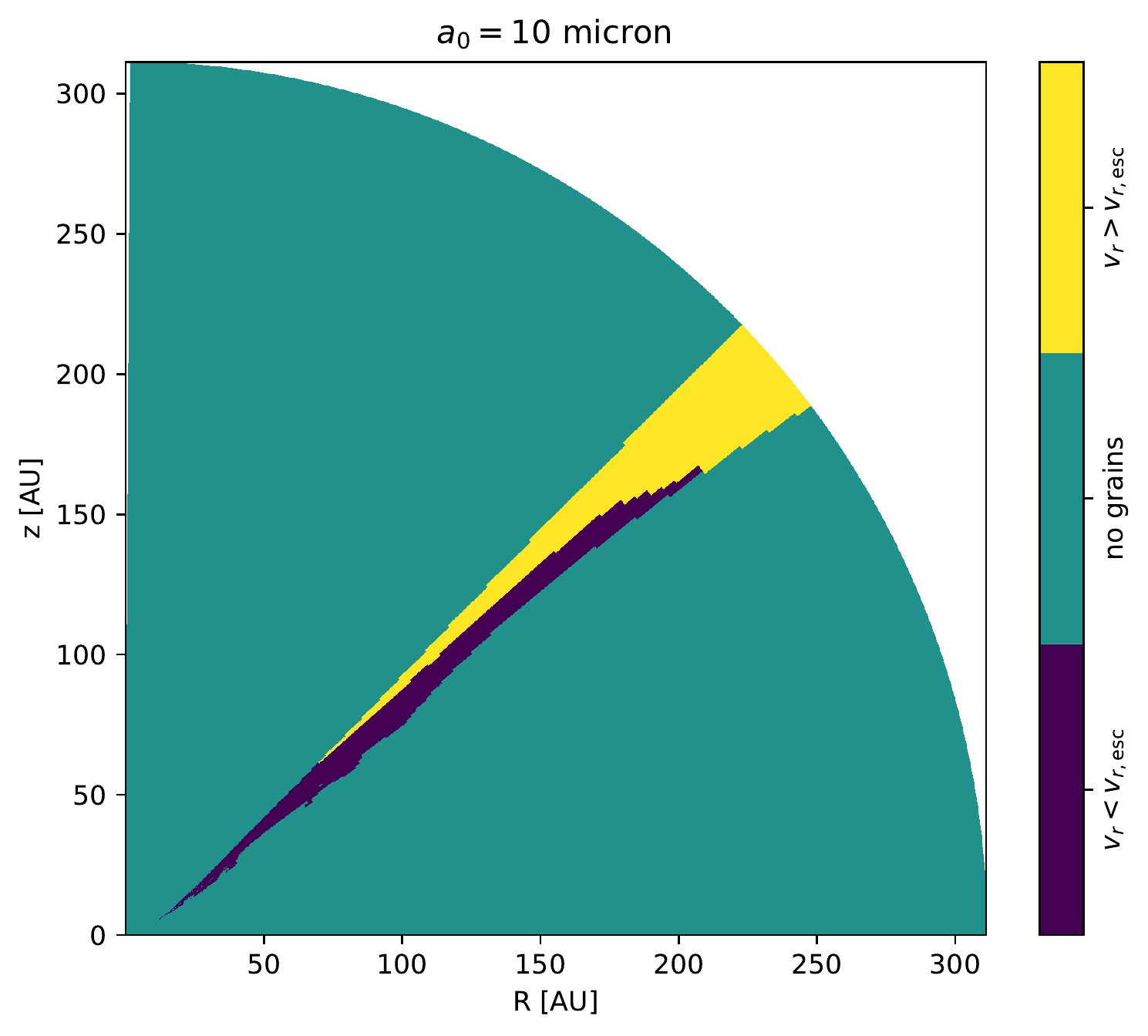}
    \caption{Acceleration of entrained grains of $a_0 = 10\,\mu$m: wherever there is at least one particle with $v_r < v_{r,\mathrm{esc}}$, the area is colored in dark blue; if all grains have reached $v_r > v_{r,\mathrm{esc}}$, it is yellow; cyan areas are not traversed by any grains of this size.
    Even for these comparatively large grains, $v_{r,\mathrm{esc}}$ is reached well within the computational domain (if at all).}
    \label{fig:vesc-region}
\end{figure}

As outlined in Sections~\ref{sec:Methods} and \ref{sec:Results}, our model was set up with no initial velocity spread, that is for all our grains, we set $\vec{v}_0 = r \, \Omega_K \, \hat{\vec{v}}_\varphi$.
Fig.~\ref{fig:vesc-region} shows that even large entrained grains reach $v_{r,\mathrm{esc}}$ well inside the computational domain (i.e. at $r \ll 300\,$AU); so we may assume that a variation of the starting velocities will not strongly affect $\max(a_0)$.

However, some turbulence is needed to vertically transport the dust particles to base of the wind; hence, we would realistically assume some spread in $\vec{v}_0$.
The main contribution to vertical mixing stems from $v_\vartheta$ which we found to be rather low in the vicinity of the base of the wind ($|v_\vartheta| \lesssim 50\,$m/s).
So in order to check the effects of varying starting velocities, we introduced a Gaussian spread of $\sigma(v_i) = 100\,$m/s in all three coordinate directions $i \in \lbrace r, \vartheta, \varphi \rbrace$.

\begin{figure}
    \centering
    \includegraphics[scale=0.55]{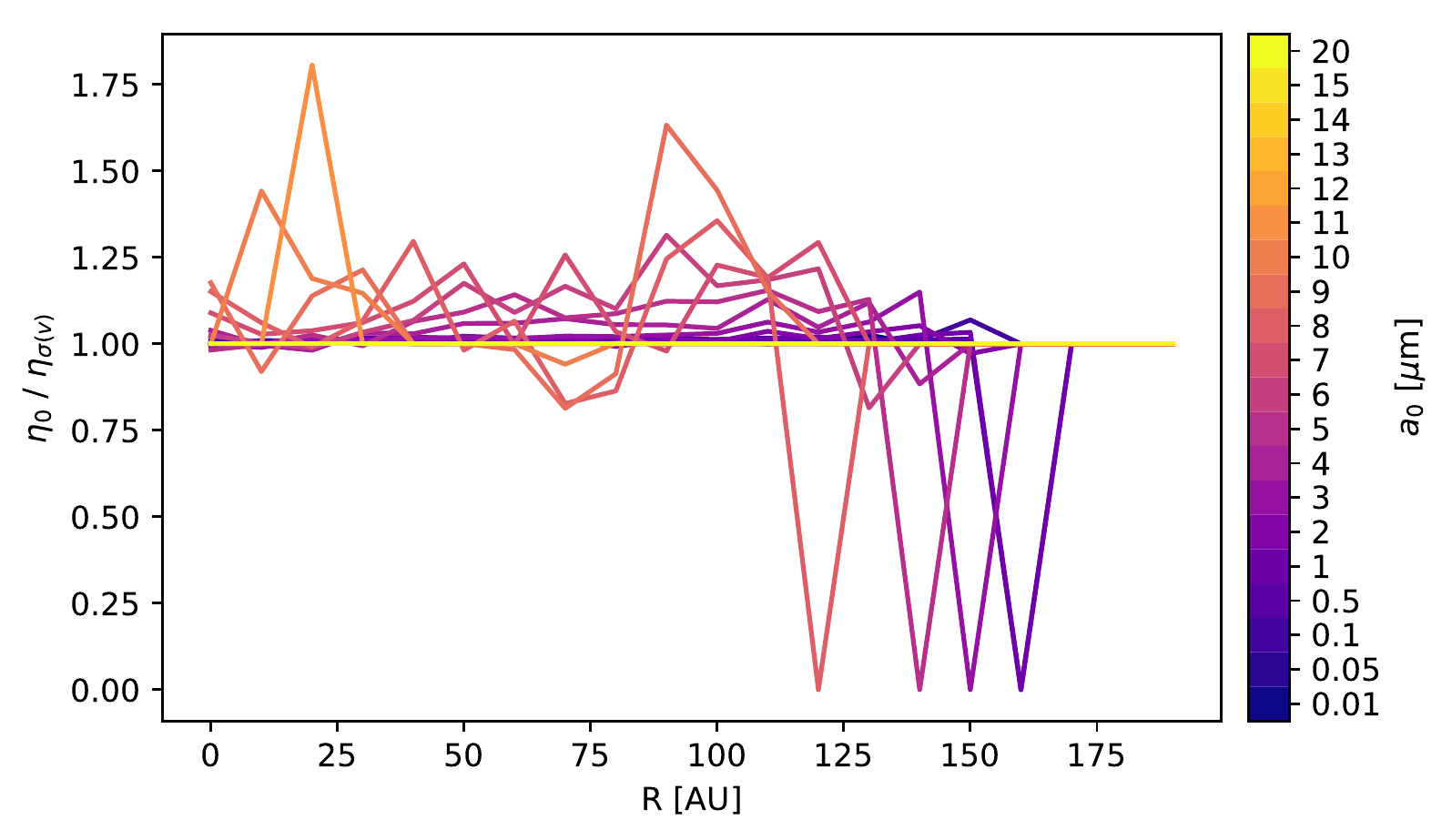}
    \caption{Ratio of the fraction of grains entrained for $\sigma(v) = 0$ (i.e. $\eta_0$) and the fraction of grains entrained for $\sigma(v) = 100\,$m/s (i.e. $\eta_{\sigma(v)}$) for bins of 10\,AU along the launching area.
    For clarity, we required at least 20 grains to be entrained for either $\eta > 0$.
    While there are distinct deviations between $\eta_0$ and $\eta_{\sigma(v)}$, they do not exhibit a clear pattern. Dips to 0 indicate regions where $\sigma(v) > 0$ allows for additional wind pick-up of some dust particles that could not have been entrained with $\sigma(v)=0$.}
    \label{fig:entrain-w-wo}
\end{figure}

As a first step, we compared entrainment ratios from along the base of the flow. We expect them to differ due to grains with a reduced (enhanced) upwards velocity being less (more) likely to be picked up by the photoevaporative wind.
In Fig.~\ref{fig:entrain-w-wo}, we see a comparison of said entrainment ratios for $\sigma(v) = 0$ (labelled $\eta_0$) and $\sigma(v) = 100\,$m/s (labelled $\eta_{\sigma(v)}$).

While we encounter some statistical variation between the two, there is no clear systematic distinction.
Grains with $a_0 \lesssim 5\,\mu$m show little variation, with $0.95 \lesssim \eta_0 / \eta_{\sigma(v)} \lesssim 1.2$; so overall, they are slightly less likely to be picked up if $\sigma(v) > 0$.
This may seem counterintuitive at first; it is probably a consequence of the combination of three individual directional, randomly positive or negative offsets to $\vec{v}_0$.
A reduced speed in either of the three coordinate directions thus may be difficult to compensate via possibly positive changes along the other two axes of motion.

Dips to 0 indicate $R$-bins where $\eta_0 = 0$ and $\eta_{\sigma(v)} > 0$; in order to avoid a series of minima produced by only very few stray grains, we have introduced a threshold of at least twenty grains to be entrained for either $\eta > 0$.

For $6 \lesssim a_0\,\mathrm{[\mu m]} \leq 11$, entrainment fractions may vary by a factor of up to 2; in addition, the peaks are more pronounced due to the lower-number statistics for larger grains (see Table~\ref{tab:counts}).

So while we may expect the dust content of the wind to vary according to the strength of the turbulent mixing, this should not be detrimental to the rest of our findings.

\begin{figure}
    \centering
    \includegraphics[scale=0.625]{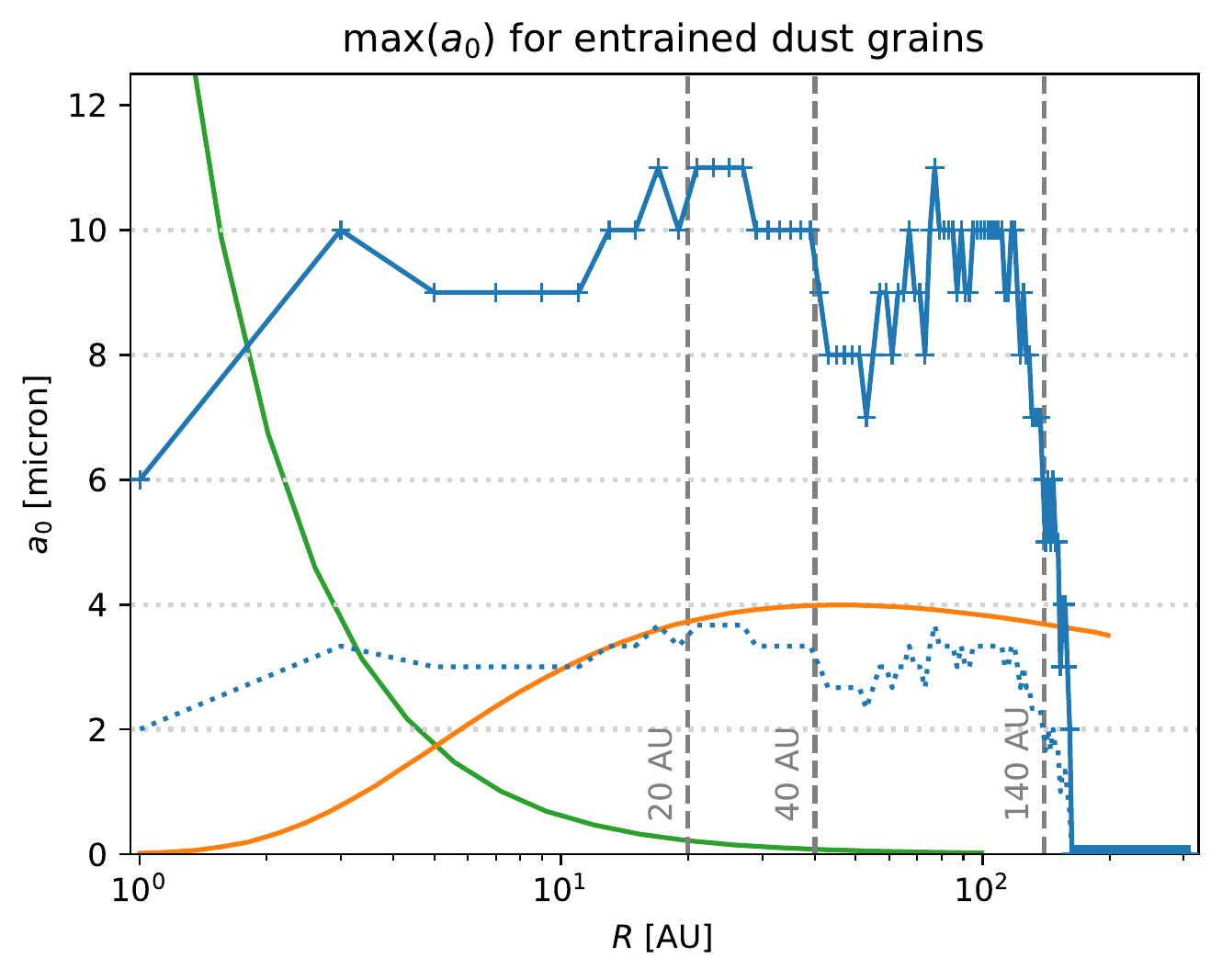}
    \caption{Maximum entrained grain size $\max(a_0)$ along the disk surface when $\sigma(v_i) = 100\,$m/s is included in the grain setup.
    The differences to Fig.~\ref{fig:max(a0)-1D}, which forgoes the velocity spread, are minor, and mostly due to numerical variations of the initial positioning.}
    \label{fig:max(a0)-1D-velsigma}
\end{figure}

\begin{figure}
    \centering
    \includegraphics[scale=0.55]{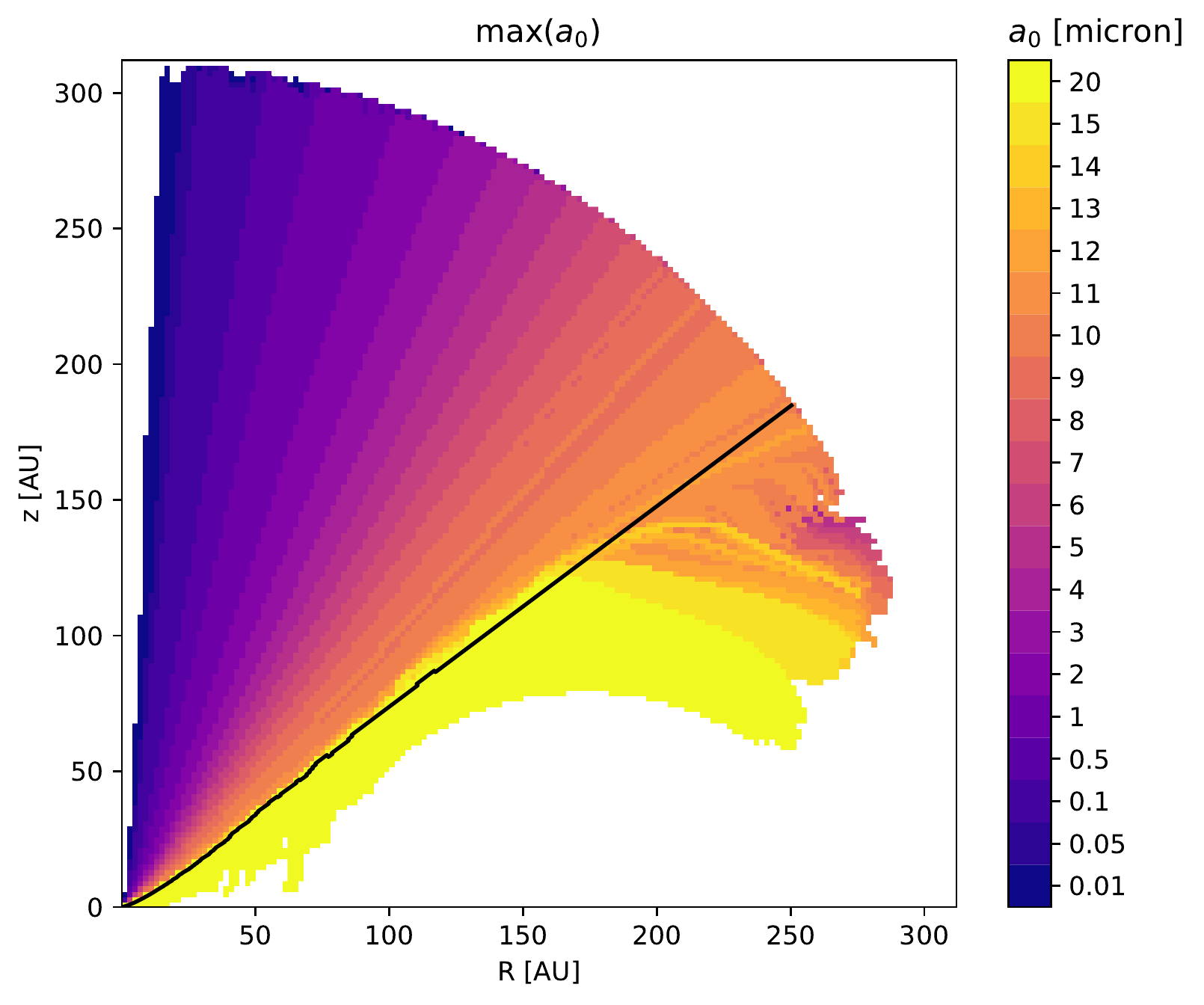}
    \caption{Maximum entrained grain size $\max(a_0)$ in 2D, to be compared to Fig.~\ref{fig:max(a0)-2D}.
    As with Figs.~\ref{fig:max(a0)-1D} and \ref{fig:max(a0)-1D-velsigma}, the differences caused by $\sigma(v_i)$ are minor at best.}
    \label{fig:max(a0)-2D-velsigma}
\end{figure}

\begin{figure}
    \centering
    \includegraphics[scale=0.55]{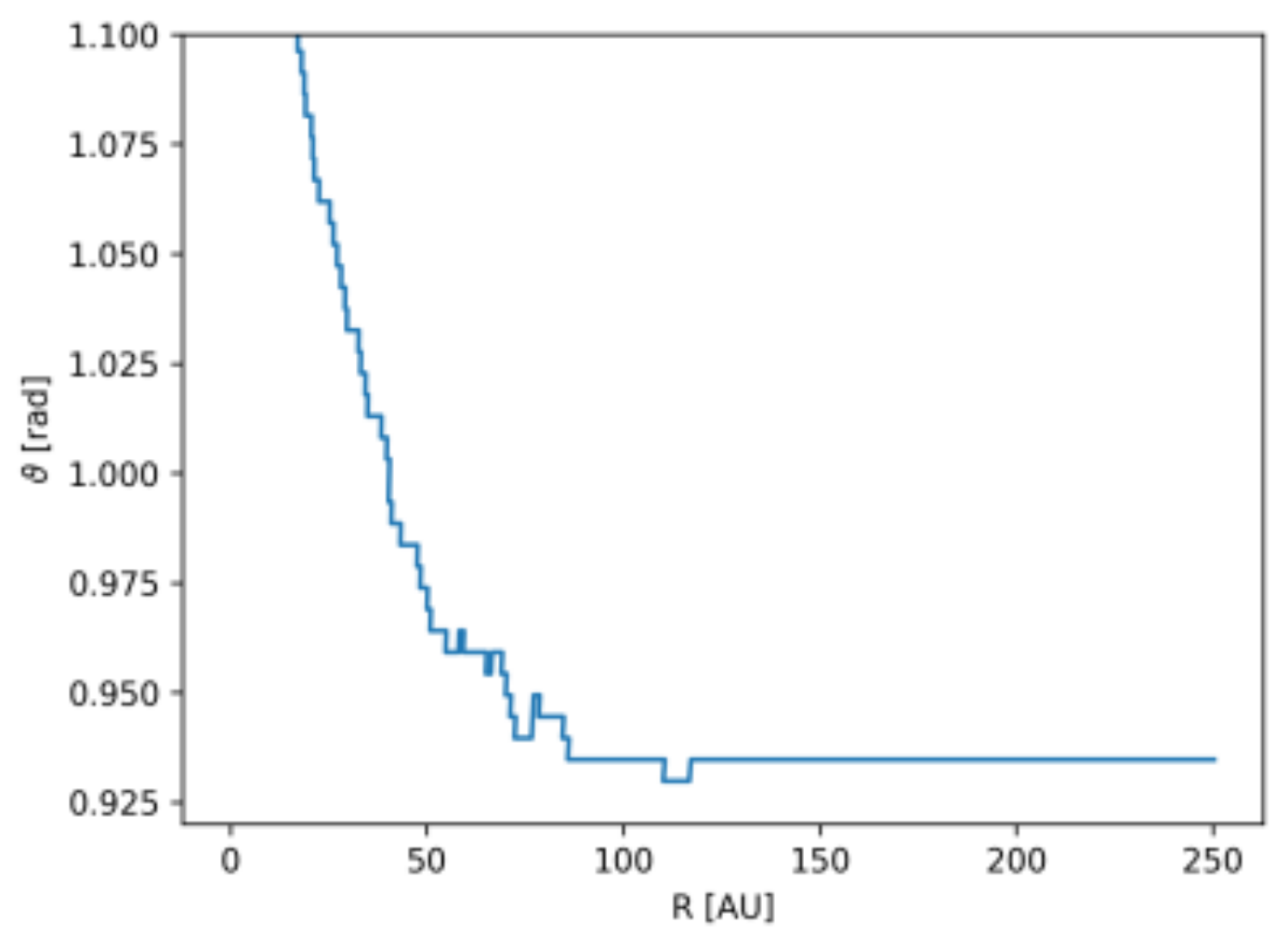}
    \caption{$\vartheta$-coordinate of the base of the photoevaporative flow. Its (slightly) craggy shape is especially pronounced around $50 \lesssim R\,[\mathrm{AU}] \lesssim 90$, which leads to small numerical artifacts in our results.}
    \label{fig:gasdisk-surface}
\end{figure}

Furthermore, the maximum entrained grain size is affected merely marginally; this can be checked when comparing Figs.~\ref{fig:max(a0)-1D-velsigma} and \ref{fig:max(a0)-2D-velsigma} to Figs.~\ref{fig:max(a0)-1D} and \ref{fig:max(a0)-2D}, respectively.
$\max(a_0)$ does vary slightly along the base of the wind; this is to be expected due to the randomized initial placement of the dust grains.
Yet the position of the peak is well-preserved -- just as for $\sigma(v)=0$, we find $\max(a_0) = 11\,\mu$m at $R \simeq 20\,$AU.
The additional local maxima stem from the numerical unevenness of the launching area, demonstrated in Fig.~\ref{fig:gasdisk-surface}.

If a grain with a velocity vector that is especially enhanced in direction of the wind motion is launched from an edge point along the base line, this may allow it to enter the wind region in contrast to another particle starting from the same location, but with a more downwards-biased velocity vector.
The 2D maps for $\max(a_0)$ in Figs.~\ref{fig:max(a0)-2D} and \ref{fig:max(a0)-2D-velsigma} are almost identical, apart from few very narrow lines from individual grains of $a_0 \geq 9\,\mu$m; it stands to reason that these are caused by the randomness in the initial conditions just described.

The fits to the parameters $c_1$ and $c_2$ of Eq.~\eqref{eq:fit-max(z)}, shown in Fig.~\ref{fig:N(R,z)-fitparam}, are almost entirely unaffected by the additional $\sigma(v_i)$; for brevity, suffice to state that both $c_1$ and $c_2$ deviate but in the third significant figure.

To summarize, while we must assume that there is a sensible amount of gas turbulence around the disk surface, this does not strongly affect our findings.


\section{Plots and parameters for all dust grain sizes}
\label{sec:app:plots-all}

We have modelled 20 distinct grain sizes, listed in Fig.~\ref{fig:dust-positions} and Table~\ref{tab:counts}; for clarity we have chosen to only show plots for three distinct sizes above.
In the following, we present the corresponding plots for all $a_0$, and argue why we have chosen exactly these to represent the full sample.

\begin{figure*}
    \centering
    \includegraphics[scale=0.235]{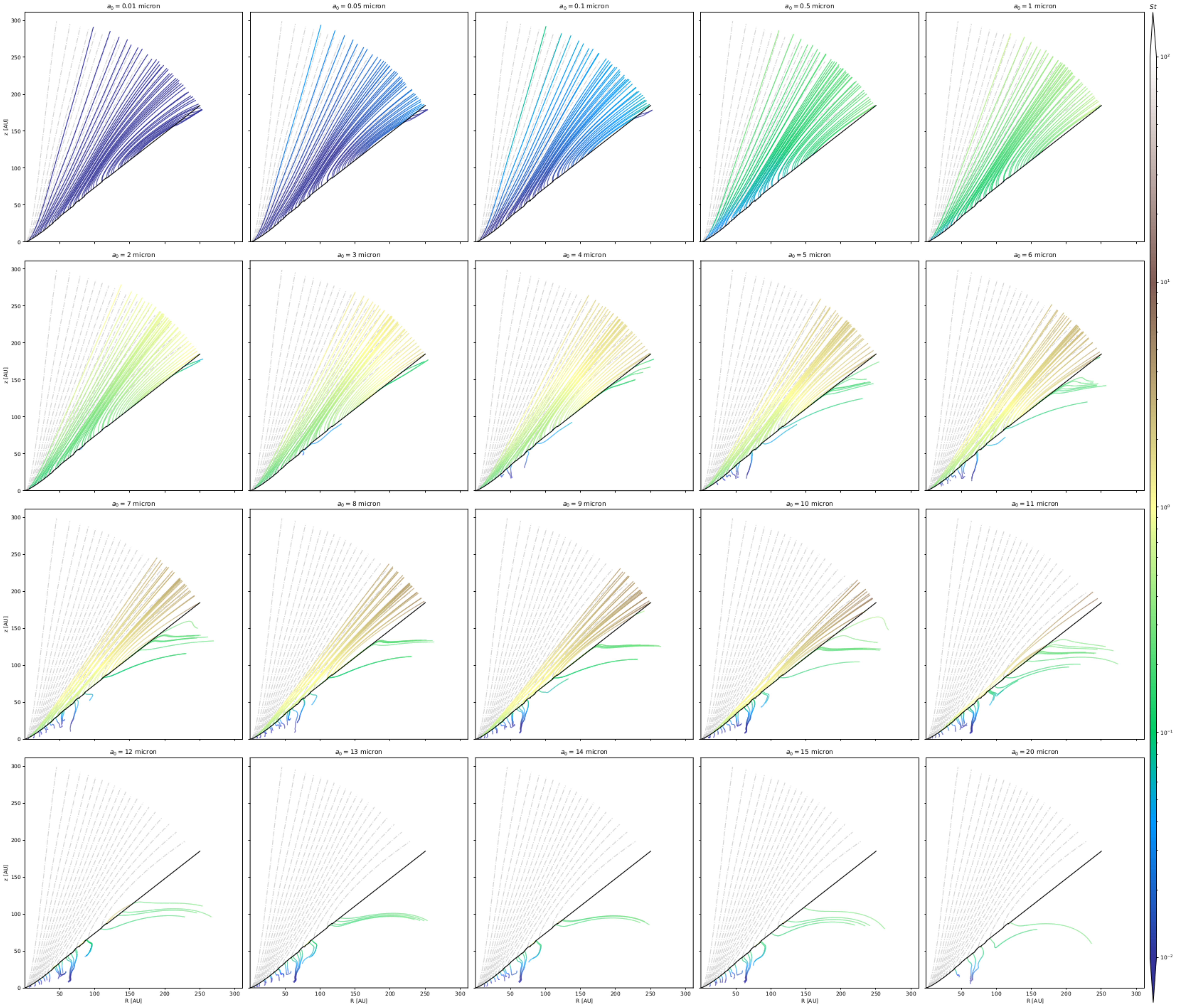}
    \caption{Full version of Fig.~\ref{fig:trajectories-samples-3}: Arbitrary selection of dust grain trajectories for all 20 $a_0$, colored by their local $St$. Wind base in black, gas streamlines in 5\% steps of $\dot{M}_\mathrm{w}$ in dash-dotted grey.}
    \label{fig:trajectories-samples-all}
\end{figure*}

An arbitrary selection of dust grain trajectories, spaced by roughly 5\,AU intervals, and entering the wind-dominated region at various points along its base is shown in Fig.~\ref{fig:trajectories-samples-all}, the full version of Fig.~\ref{fig:trajectories-samples-3}. The individual trajectories are colored according to their local Stokes number $St$.

There are four main scenarios to be found here:

Firstly, full wind entrainment.
This applies if $St \ll 1$ throughout the trajectories, that is for $a_0 \lesssim 0.1\,\mu$m.
Visually, the trajectories shown for these grain sizes are mostly blue, indicating $St < 0.1$.
As discussed above, these dust particles follow the gas motion very closely; this causes grains to fall back below the base of the wind along the gas streamlines, that is for $R \gtrsim 180\,$AU.
The most massive grains of this group have sizes $a_0 = 0.1\,\mu$m, which we have chosen for the plots in Section~\ref{sec:Results}.
    
Secondly, slow decoupling from the gas flow.
For $0.5 \lesssim a_0\,[\mu\mathrm{m}] \lesssim 5$, the grains are picked up with $St \ll 1$. Yet while the particles are blown out and hence within $r < 300\,$AU, their $St$ approaches 1.
Thus the trajectories decouple from the gas; high above the disk surface, they fall below the gas streamlines, whereas close to it, they deviate upwardly, leading to a more radial outflow in all cases.
Grain fallback occurs for $R \gtrsim 170\,AU$.
$a_0 = 4\,\mu$m represents a grain size from this interval for which $St \rightarrow 1$ is readily apparent.
    
Thirdly, quick decoupling.
For $6 \lesssim a_0\,[\mu\mathrm{m}] \leq 11$, the interval during which the dust grains follow the gas stream is shorter, and the grains reach higher $St$ while being blown out.
For $R \gtrsim 160\,AU$, we observe particles falling back below the base of the flow because the wind cannot provide enough momentum.
The almost-largest grain size in this group is $a_0 = 10\,\mu$m ($11\,\mu$m grains are more sparse in the wind region).

Fourthly, no wind pick-up.
For $a_0 > 11\,\mu$m, the grains are too heavy to be lifted up by the wind; even if they reach the disk surface, they fall back below it and successively follow the gas streams in the disk because of the comparably high densities there.
Since we find no entrainment for these grains, we have omitted this group from the plots in Section~\ref{sec:Results}.

\begin{figure*}
    \centering
    \includegraphics[scale=0.24]{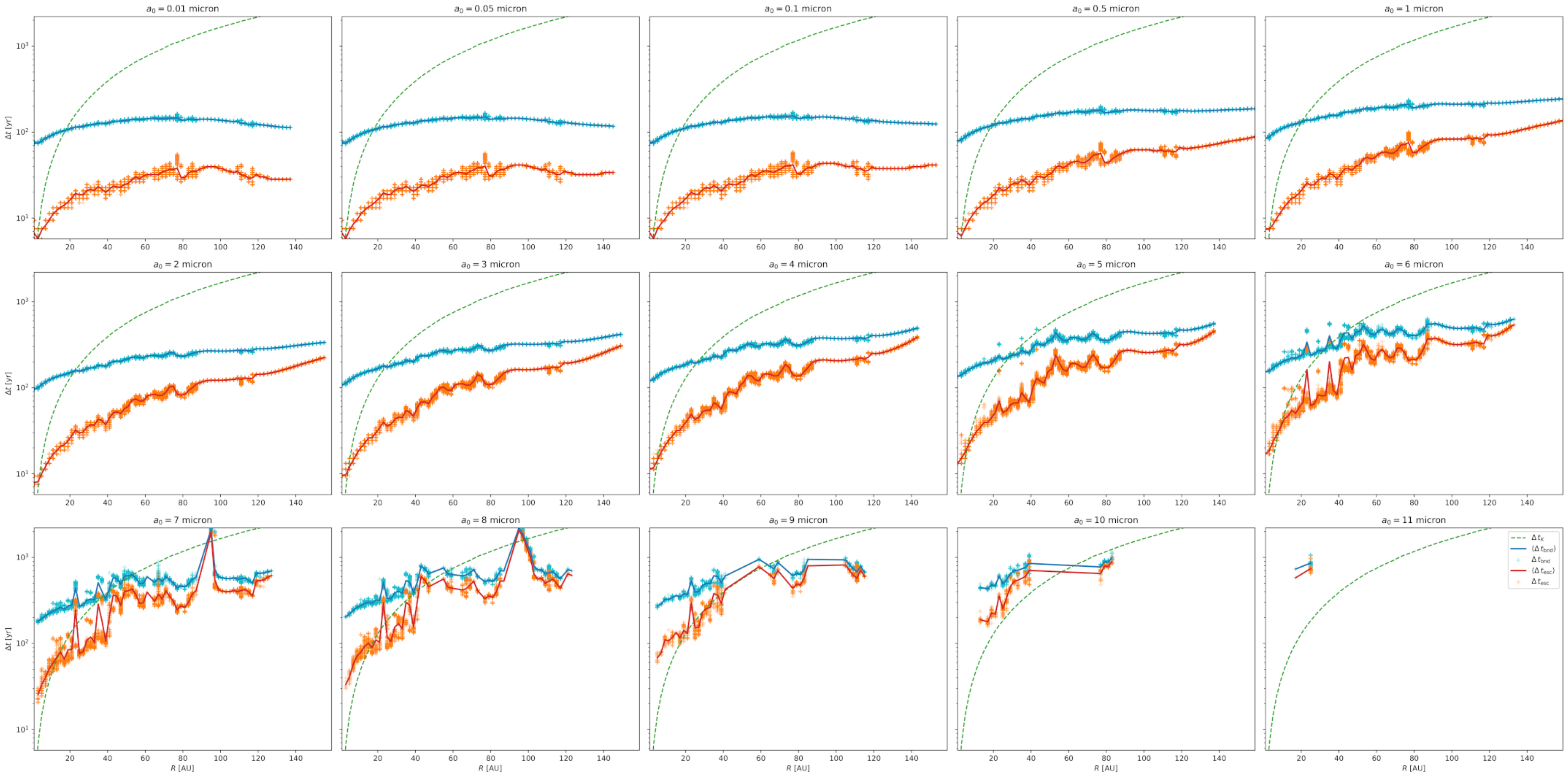}
    \caption{Times needed to fully blow out dust particles from their starting position along the base of the wind to the domain boundary (i.e. $\Delta t_\mathrm{bnd}$) and to accelerate them to $v_{r,\mathrm{esc}}$ (i.e. $\Delta t_\mathrm{esc}$).
    Individual data points in cyan and orange and mean values in blue and red, respectively.
    The rasterization of the former points results from a time-discrete particle tracking and a binning in $R$-direction; a raster point may therefore represent multiple data points.
    Keplerian orbital times at the base of the wind are included as dashed green lines.}
    \label{fig:timescales-vel-all}
\end{figure*}

Expanding on Fig.~\ref{fig:timescales-vel-3}, Fig.~\ref{fig:timescales-vel-all} shows the times it takes the dust grains to reach $r \gtrsim 300\,$AU after wind pick-up -- labelled $\Delta t_\mathrm{bnd}$ (data points in cyan, mean in blue) -- and the (much lower) times for acceleration to the escape velocity $v_{r,\mathrm{esc}}$ -- labelled $\Delta t_\mathrm{esc}$ (data points in orange, mean in red).\footnote{As outlined in Section~\ref{sec:Results}, discrete output times and binning along $R$ lead to a rasterization of the individual data points for $\Delta t_\mathrm{bnd}$ and $\Delta t_\mathrm{esc}$; hence most plotted points represent much more than one data point.}

Choosing the same size categories as above, with the panels for $a_0 > 11\,\mu$m omitted since such grains are not entrained by the photoevaporative flow, we find the following:

Firstly, very small dust grains.
These are accelerated very strongly when entering the wind close to the star, resulting in low $\Delta t_\mathrm{bnd}$ and even lower $\Delta t_\mathrm{esc}$; at higher $r$, the timescales increase.
At $R \approx 80\,$AU, $\Delta t_\mathrm{bnd}$ peaks and starts falling off again because grains launching from further out have less distance to cover to the computational boundary at $r \simeq 300\,$AU.
Entrainment occurs for $R \lesssim 140...160\,$AU, depending on $a_0$; the according time frames span $70 \lesssim \Delta t_\mathrm{bnd}\,[\mathrm{yr}] \lesssim 170$ and $5 \lesssim \Delta t_\mathrm{esc}\,[\mathrm{yr}] \lesssim 50$.
The timescale of the particle motion in the wind is smaller than the Keplerian orbital timescale for grains launched from $R \gtrsim 20\,$AU ($\Delta t_\mathrm{bnd}$) or $R \gtrsim 1\,$AU ($\Delta t_\mathrm{esc}$), meaning that the motion in the wind dominates the dynamic evolution.
    
Secondly, small grains.
They show a monotonic increase of $\Delta t_\mathrm{bnd}$ and $\Delta t_\mathrm{esc}$ with $R$; the peak we have seen for smaller $a_0$ disappears.
Hence, the farther out a dust grain is picked up by the XEUV wind, the longer it takes to be blown out of the computational domain and to be accelerated to $v_{r,\mathrm{esc}}$; while this may appear counter-intuitive at first, it merely illustrates that the gas flow is much stronger closer to the star.
The grain transport happens on a timescale of a few yr to a few $10^2\,$yr; the Keplerian motion dominates the dynamic timescale for $R \gtrsim 30\,$AU ($\Delta t_\mathrm{bnd}$) or $R \gtrsim 5\,$AU ($\Delta t_\mathrm{esc}$).

Thirdly, medium-sized grains.
As for the small grains, we see a mostly monotonic relation between $R$ and $\Delta t_\mathrm{bnd}$.
However, the graph exhibits very distinct peaks for $a_0\,[\mu\mathrm{m}] \in \lbrace 8; 9 \rbrace$, which we have not commented on in Sections~\ref{sec:Methods} and \ref{sec:Discussion} because they are numerical artifacts caused by the craggy launching area (see Section~\ref{sec:Results} and Appendix~\ref{sec:app:vel-sigma}), and interestingly disappear in the setup including an initial velocity spread.\footnote{Note that the $\Delta t_\mathrm{bnd}$ data points here have been retrieved for fully entrained grains which have left the computational domain within the computational time frame $\Delta t_\mathrm{sim}$. Thus, the data points almost reaching our $\Delta t_\mathrm{sim}$ do not indicate that the latter is too short to capture their full trajectory.}
Apart from that, we find $t_\mathrm{dyn} < \Delta t_\mathrm{bnd}$ for $R \lesssim 60\,$AU and an almost identical value for $\Delta t_\mathrm{esc}$, indicating that speed-up is slower for heavier grains, as would be expected.

For $R \rightarrow 0$, we find $\Delta t_\mathrm{esc} \rightarrow 10\,$yr for $a_0 \lesssim 9\,\mu$m.
This is another indicator that the photoevaporative flow is strongest close to the star, where it is most effective at accelerating the dust grains.

Last, Fig.~\ref{fig:N(R,z)-all} shows the areas occupied by the dust grains, and the according fits for $\left. \max(z) \right|_R$.
The fit parameters are also listed in Table~\ref{tab:(R,max(z))}, with $c_1$ and $c_2$ as defined in Eq.~\eqref{eq:fit-max(z)}.

\begin{figure*}
    \centering
    \includegraphics[scale=0.24]{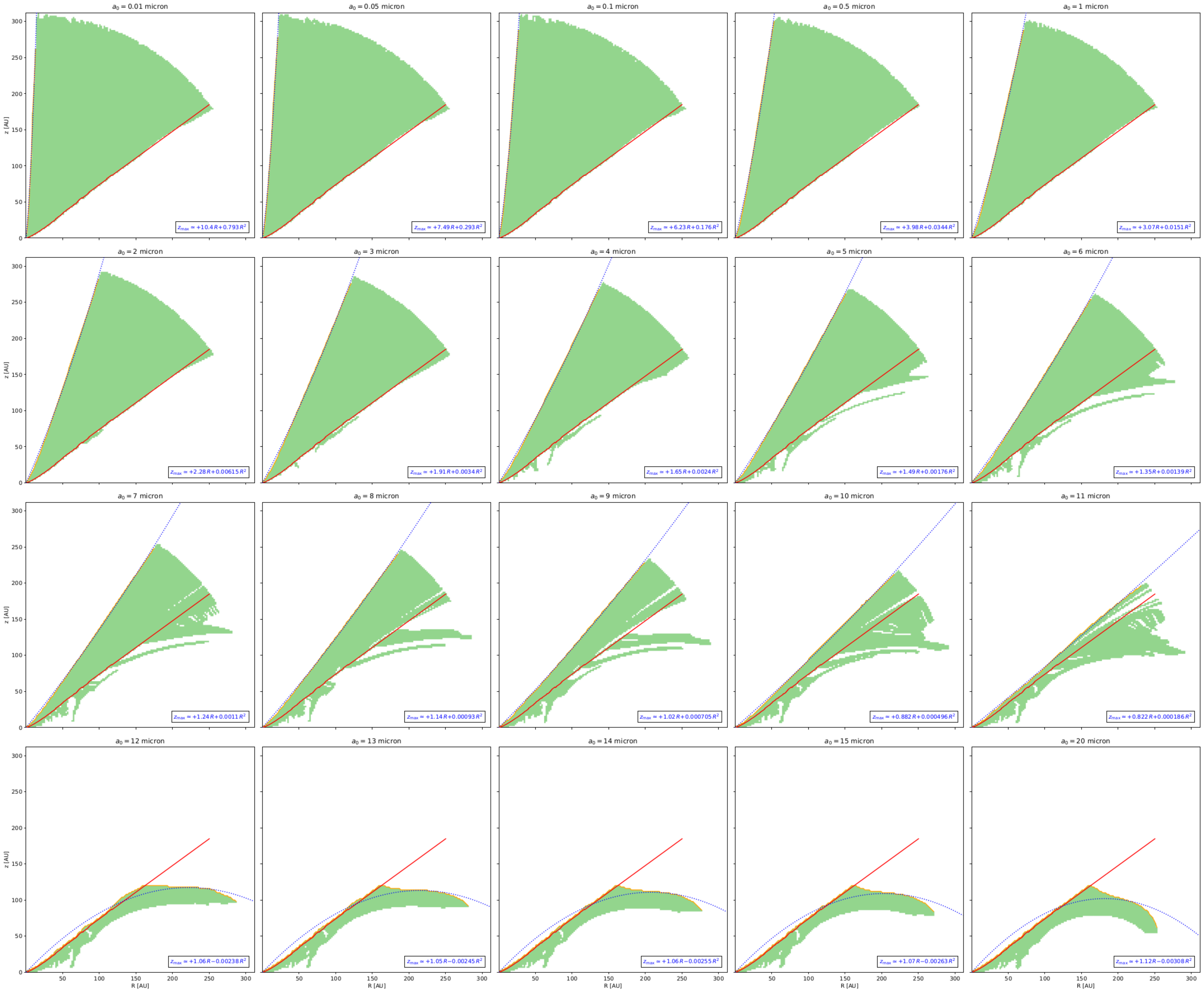}
    \caption{Full version of Fig.~\ref{fig:N(R,z)-3}: Dust population in the wind for all $a_0$. We see a clear correlation between $R$ and $\max(z)$ in the individual plots, fitted with dashed blue lines. The parameters given in the plots also listed in Table~\ref{tab:(R,max(z))}. Base of the wind in red, non-fitted population boundaries in orange.}
    \label{fig:N(R,z)-all}
\end{figure*}

\begin{table}
    \centering
    \caption{Fit parameters and standard errors for $\left. \max(z) \right|_R$ for entrained grains, retrieved as shown in Fig.~\ref{fig:N(R,z)-3}, for the fit provided in Eq.~\eqref{eq:fit-max(z)}.}
    \label{tab:(R,max(z))}
    \begin{tabular}{*{5}{r}}
        \multicolumn{1}{l}{$a_0$[$\mu$m]} & \multicolumn{1}{l}{$c_1$} & \multicolumn{1}{l}{$\sigma(c_1)$} & \multicolumn{1}{l}{$c_2$} & \multicolumn{1}{l}{$\sigma(c_2)$} \\ \hline
        $0.01$ & $10.4$ & $1.80$ & $7.93\cdot 10^{-1}$ & $1.67\cdot 10^{-1}$ \\
        $0.05$ & $7.49$ & $5.96\cdot 10^{-1}$ & $2.93\cdot 10^{-1}$ & $3.51\cdot 10^{-2}$ \\
        $0.1$ & $6.23$ & $3.40\cdot 10^{-1}$ & $1.76\cdot 10^{-1}$ & $1.57\cdot 10^{-2}$ \\
        $0.5$ & $3.98$ & $1.10\cdot 10^{-1}$ & $3.44\cdot 10^{-2}$ & $2.63\cdot 10^{-3}$ \\
        $1$ & $3.07$ & $5.94\cdot 10^{-2}$ & $1.51\cdot 10^{-2}$ & $1.07\cdot 10^{-3}$ \\
        $2$ & $2.28$ & $3.08\cdot 10^{-2}$ & $6.15\cdot 10^{-3}$ & $3.97\cdot 10^{-4}$ \\
        $3$ & $1.91$ & $2.01\cdot 10^{-2}$ & $3.40\cdot 10^{-3}$ & $2.13\cdot 10^{-4}$ \\
        $4$ & $1.65$ & $1.51\cdot 10^{-2}$ & $2.40\cdot 10^{-3}$ & $1.41\cdot 10^{-4}$ \\
        $5$ & $1.49$ & $1.14\cdot 10^{-2}$ & $1.76\cdot 10^{-3}$ & $9.65\cdot 10^{-5}$ \\
        $6$ & $1.35$ & $9.47\cdot 10^{-3}$ & $1.39\cdot 10^{-3}$ & $7.45\cdot 10^{-5}$ \\
        $7$ & $1.24$ & $8.82\cdot 10^{-3}$ & $1.10\cdot 10^{-3}$ & $6.47\cdot 10^{-5}$ \\
        $8$ & $1.14$ & $7.20\cdot 10^{-3}$ & $9.30\cdot 10^{-4}$ & $5.00\cdot 10^{-5}$ \\
        $9$ & $1.02$ & $5.43\cdot 10^{-3}$ & $7.05\cdot 10^{-4}$ & $3.50\cdot 10^{-5}$ \\
        $10$ & $0.882$ & $6.19\cdot 10^{-3}$ & $4.96\cdot 10^{-4}$ & $3.66\cdot 10^{-5}$ \\
        $11$ & $0.822$ & $5.65\cdot 10^{-3}$ & $1.86\cdot 10^{-4}$ & $3.15\cdot 10^{-5}$ \\
        $12$ & $1.06$ & $1.39\cdot 10^{-2}$ & $-2.38\cdot 10^{-3}$ & $6.21\cdot 10^{-5}$ \\
        $13$ & $1.05$ & $1.42\cdot 10^{-2}$ & $-2.45\cdot 10^{-3}$ & $6.52\cdot 10^{-5}$ \\
        $14$ & $1.06$ & $1.55\cdot 10^{-2}$ & $-2.55\cdot 10^{-3}$ & $7.18\cdot 10^{-5}$ \\
        $15$ & $1.07$ & $1.67\cdot 10^{-2}$ & $-2.63\cdot 10^{-3}$ & $7.92\cdot 10^{-5}$ \\
        $20$ & $1.12$ & $2.34\cdot 10^{-2}$ & $-3.08\cdot 10^{-3}$ & $1.19\cdot 10^{-4}$
    \end{tabular}
\end{table}

Firstly, as claimed in Section~\ref{sec:Results}, these plots illustrate that all $a_0 < \max(a_0)$ populate the regions occupied by the $\max(a_0)$ shown in Fig.~\ref{fig:max(a0)-2D}.
Secondly, the unpopulated regions found in Fig.~\ref{fig:N(R,z)-all} are once again caused by the slight bumpiness of the base of the flow; additional test runs with a much increased spatial resolution (i.e. 1000 particles per 1\,AU along the launching plane) were used to confirm this -- the plots for them look very similar (and have therefore not been included here).

\end{appendix}

\end{document}